\documentclass[pdflatex,sn-mathphys]{sn-jnl}

\jyear{2021}
\theoremstyle{thmstyleone}

\theoremstyle{thmstyletwo}%

\theoremstyle{thmstylethree}%
\usepackage{graphicx}
\usepackage{amsmath}
\usepackage{amssymb}
\usepackage{multirow}
\usepackage{array}
\usepackage{makecell}
\NewExpandableDocumentCommand\XGap{m}{\noalign{\vskip #1}}
\NewExpandableDocumentCommand\Gap{}{\XGap{3pt}}
\newif\iffinal
\finaltrue 

\iffinal
\else
    \usepackage{xcolor}
    \usepackage[normalem]{ulem}
    \newcommand{\added}[1]{{\color{blue} #1}} 
    \newcommand\deleted{\bgroup\markoverwith{\textcolor{blue}{\rule[0.5ex]{2pt}{0.8pt}}}\ULon}
\fi

\iffinal
    \newcommand{\added}[1]{#1} 
    \newcommand\deleted[1]{}    
\fi

\begin{document}

\title[Generative Language Models]{Generative Language Models Potential for Requirement Engineering Applications: Insights into Current Strengths and Limitations}

 \author*[1,2]{\fnm{Summra} \sur{Saleem}}\email{summra.saleem@dfki.de}
\equalcont{These authors contributed equally to this work.}

\author*[2]{\fnm{Muhammad Nabeel} \sur{Asim}}\email{Muhammad\_Nabeel.Asim@dfki.de}
 \equalcont{These authors contributed equally to this work.}

\author[2]{\fnm{Ludger} \sur{Van Elst}}\email{elst@dfki.de}
\author[1,2]{\fnm{Andreas} \sur{Dengel}}\email{andreas.dengel@dfki.de}

 \affil[1]{\orgdiv{Department of Computer Science}, \orgname{Rhineland-Palatinate Technical University of Kaiserslautern-Landau}, \orgaddress{\city{Kaiserslautern}, \postcode{67663},  \country{Germany}}}
 
\affil[2]{\orgname{German Research Center for Artificial Intelligence GmbH}, \orgaddress{\city{Kaiserslautern}, \postcode{67663},  \country{Germany}}}


\abstract{Traditional language models have been extensively evaluated for software engineering domain, however \added{the} potential of ChatGPT and Gemini have not been fully explored. To fulfill this gap, \added{the} paper in hand presents a comprehensive case study to investigate the potential of both language models for development of diverse types of requirement engineering applications. It deeply explores impact of varying levels of expert knowledge prompts on the prediction accuracies of both language models. Across 4 different \deleted{requirement engineering tasks} public benchmark datasets \added{of requirement engineering tasks}, it compares performance of both language models with existing task specific machine/deep learning predictors and traditional language models. Specifically, the paper utilizes 4 benchmark datasets; Pure (7, 445 samples, requirements extraction),PROMISE (622 samples, requirements classification), REQuestA (300 question answer (QA) pairs) and Aerospace datasets (6347 words, requirements NER tagging).  \deleted{A thorough experimentation reveals} \added{Our experiments reveal} that, in comparison to ChatGPT, Gemini requires more careful prompt engineering to provide accurate predictions. Moreover, across requirement extraction benchmark dataset \added{the} state-of-the-art F1-score is 0.86 while ChatGPT and Gemini achieved 0.76 and 0.77, respectively. \added{The} State-of-the-art F1-score \deleted{onn} \added{on} requirements classification dataset is 0.96 and both language models \deleted{managed to produce} \deleted{yielded} 0.78. In name entity recognition (NER) task \added{the} state-of-the-art F1-score is 0.92 and ChatGPT managed to produce 0.36, and Gemini 0.25. Similarly, across question answering dataset \added{the} state-of-the-art F1-score is 0.90 and ChatGPT and Gemini managed to produce 0.91 and 0.88 respectively. \added{Our experiments show that Gemini requires more precise prompt engineering than ChatGPT. Except for question-answering, both models under-perform compared to current state-of-the-art predictors across other tasks.}}




\maketitle

\section{Introduction}
In recent years, the world has witnessed a remarkable surge for development of diverse and impact-full computer-aided applications \citep{palmquist2020applications, oza2021machine}. The \deleted{driving force behind}  \added{primary motivation behind the} development of these applications \deleted{stems from an ardent desire} \added{is} to automate time-consuming and labor-intensive tasks by utilizing cutting-edge Artificial Intelligence (AI) algorithms \citep{castiglioni2021ai, reddy2020governance}. In software systems domain, to facilitate development of large softwares, researchers have  proposed numerous software development models such as waterfall model \citep{petersen2009waterfall}, v model \citep{ruparelia2010software}, spiral model \citep{boehm1988spiral} and agile model \citep{abrahamsson2017agile}. In all these models requirement engineering is a fundamental and essential \deleted{task}\added{process comprising several tasks} \citep{laplante2022requirements}. \deleted{such as, requirements extraction is the process of structuring insight of user expectations from a variety of sources such as user manuals, customer feedback, emails, interviews and software requirements specification (SRS)} \added{One critical task is requirements extraction, which involves identification of stakeholders needs from heterogeneous collection of documents such as user manuals, customer feedback, emails, interviews.} \citep{pandey2010effective}. \added{Another, important task is} requirements classification \deleted{enables effective categorization of software}\added{into functional and non-functional} features which aids in \added{the} identification and mitigation of potential risks that may hinder \added{the} project success \citep{saleem2023fnreq, hey2020norbert, kurtanovic2017automatically}. Furthermore, in the context of test case generation, NER tagging is key task for extracting actors and actions \citep{mahalakshmi2018named}. Similarly, requirements-based question answering system provides accurate and relevant answers to questions based on specific requirements. This type of system is particularly valuable in various contexts which require precise information to ensure software quality. 

In \added{the} aforementioned applications, early predictive pipelines were designed using traditional machine learning (ML) algorithms \citep{nagpal2021comprehensive, althanoon2021supporting, rahimi2020ensemble, haque2019non, toth2019comparative} but had performance limitations. The invention of deep learning algorithms opened the doors of \added{a} new era for these applications with significant performance boost \citep{navarro2017towards}. Despite deep learning advancement, sub-optimal representation learning methods remained \added{a} bottleneck for applications performance  \citep{althanoon2021supporting, rahimi2020ensemble, toth2019comparative}. The need of representation learning arises due to inherit \deleted{dependencyof} \added{dependency of} machine and deep learning-based methods on statistical vectors \citep{liu2023representation, scholkopf2021toward}. To address this need, initially researchers extensively utilized one hot vector \citep{pham2018exploiting, xiong2020distributed} and \added{(Term Frequency-Inverse Document Frequency)} TFIDF  \citep{qin2016novel, liu2018research} representation learning methods. Despite their widespread use, both methods have significant shortcomings. One hot vector lacks ability to capture semantic relations between different words \citep{selva2021review}  and TFIDF representation generates sparse \deleted{matrix} \added{matrices} \citep{qin2016novel}.  Apart from representation learning, another obstacle of deep learning algorithms was their dependency on extensive annotated data \citep{najafabadi2015deep}. Many research studies \citep{kraljevic2021multi} \deleted{has} \added{have} concluded that deep learning algorithms produce superior performance with larger training data but struggle to achieve comparable performance on smaller dataset\added{s} \citep{shrestha2019review}. However, \added{generating} large annotated datasets is a resource intensive task and requires significant cost and human power \citep{hendrycks2021cuad, kraljevic2021multi}. 

The groundbreaking emergence of transfer learning empowered deep learning algorithms to excel even with limited datasets \citep{yang2020transfer, ruder2019transfer}. The domain of Natural Language Processing (NLP), leveraged word embeddings to implement the concept of transfer leaning \citep{wang2019evaluating, wang2020survey}\deleted{,} to learn representation from large un-annotated textual data \citep{azunre2021transfer}. The prime objective of word embedding methods is to generate semantic rich statistical vectors of words where similar words are positioned closer together and dissimilar words are located farther apart in vector space \citep{mehmood2023passion}. Specifically, word embeddings addressed the challenges of large training data by facilitating pre-trained word embeddings and offered more comprehensive representations as compared to one hot vector and TFIDF based representation \citep{qin2016novel, liu2018research}. However, pre-trained word embeddings provide  pre-trained weights only at embedding layer  \citep{mehmood2023passion} and remaining layers of deep learning model are initialized with random weights. Consequently, the effectiveness of pre-trained word embeddings inspired researchers to develop more generalized models capable of providing pre-trained weights for all layers. This led to  groundbreaking emergence of language model; a major breakthrough in advancement of NLP domain \citep{jansen2023employing}. 
\begin{figure*}[]
 \centering 
 \includegraphics[width=1\textwidth]{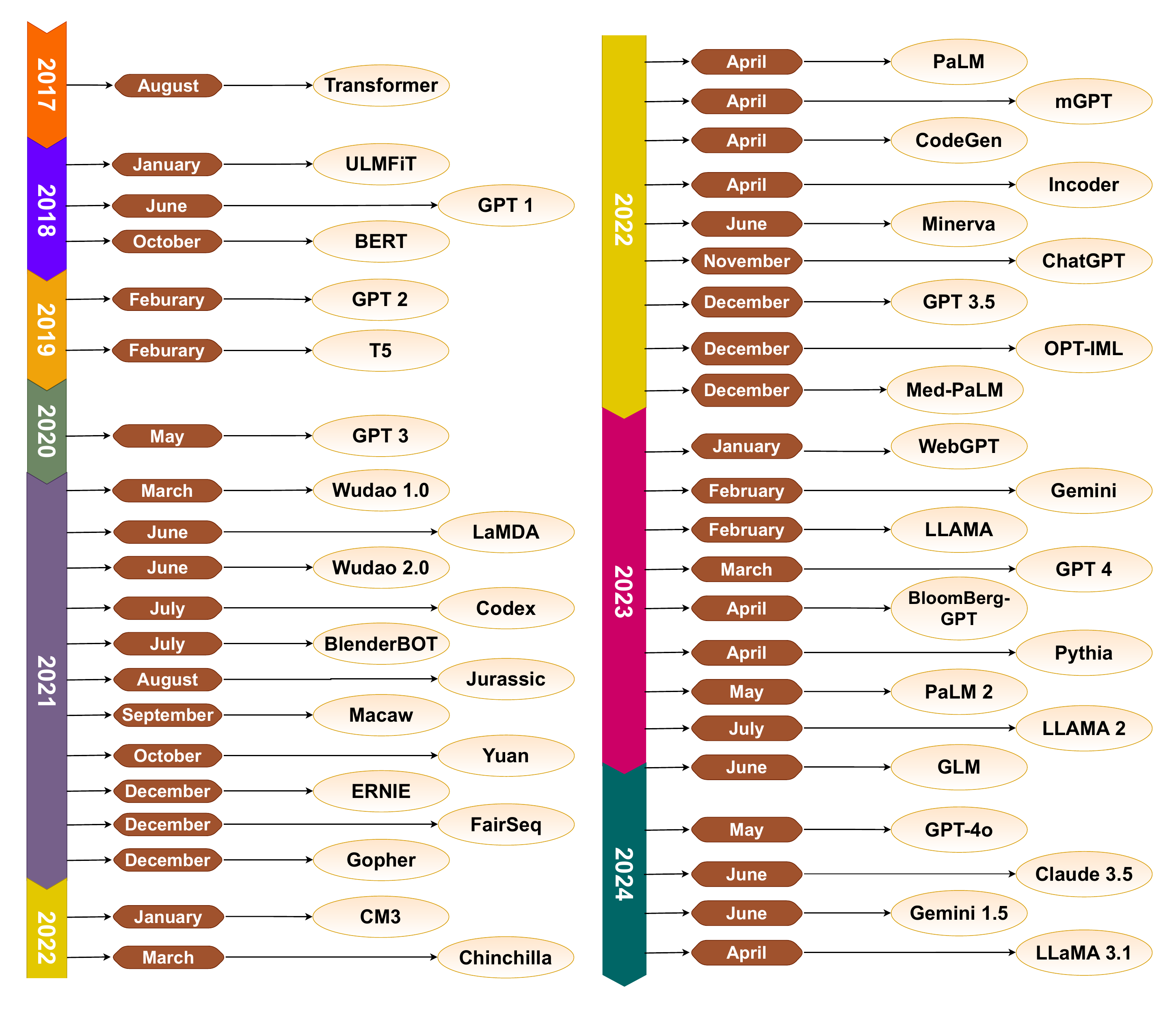}
 \caption{Landscape of language models evolution}
\label{lm-table}
\end{figure*}
\newline
\deleted{It can be seen in Table \ref{lm-table},} \deleted{o}\added{O}ver the time plethora of language models have been developed \added{(Figure \ref{lm-table})}. \added{The} \deleted{M}\added{m}otivation behind development of each new model was to leverage large unlabeled corpora to deeply understand the semantic relations of words within a certain text segment. Overall, based on working paradigm and real-world applications, existing language models can be divided into two distinct groups: those relying on fine-tuning and those requiring  prompt engineering. Extensive research has been conducted to evaluate and compare performance of language models of first category across a wide range of requirement engineering applications namely; requirements extraction \citep{ivanov2022extracting}, classification \citep{saleem2023fnreq}, duplicate requirement identification \citep{lahmerautomating}, requirements reuse \citep{sasmito2023challenges, arora2023sustainable}, NER tagging \citep{malik2021named}, question answering  \citep{ezzini2023ai} and sentiment analysis of software's reviews \citep{vijayvargiya2023software}. \deleted{Second category based} \added{Contrarily,}  language models \added{in second category} such as ChatGPT and Gemini are being extensively utilized for text re-phrasing \citep{amini4514430you, neto2021split}, data generation, healthcare \citep{ sallam2023utility, shah2023creation, muftic2023exploring}, finance \citep{li2023large, dowling2023chatgpt, rane2023role}, education \citep{kasneci2023chatgpt} and opinion mining \citep{wang2017generative} etc. However, only a limited number of research studies \citep{yeow2024automated, wu2024chatgpt, hamdi2023prompt, alter2024could}  has been performed to explore their potential across software engineering tasks. Following subsection comprehensively outlines research contributions of this study.

\subsection{Research contributions}
This research study aims to provide a thorough assessment of the ChatGPT and Gemini's capabilities in the realm of requirement engineering applications.  With an aim to deeply explore these language models, contributions of this paper are \deleted{manifold}\added{following}: 
\begin{enumerate}
    \item  \deleted{It}\added{This study} explores whether latest generative language models (ChatGPT, Gemini) \deleted{has}\added{have} potential to facilitate requirement engineering processes.
    \item  \deleted{It}\added{This study} investigates whether generative language models achieve better performance with generic prompts or prompts enriched with domain-specific knowledge. To perform this analysis, it designs three distinct types prompts  by incorporating varying levels of domain-specific knowledge, for four different requirement engineering tasks.
    \item \deleted{It}\added{This study} conducts experiments on benchmark datasets of four distinct requirements engineering tasks using three different types of prompts. It compares the performance of language models across these prompt types and presents a detailed analysis of incorrect predictions for each task.
    \item \deleted{It performs large scale comparison between ChatGPT and Gemini across 4 different requirement engineering tasks. Primarily, requirement engineering tasks fall under three distinct paradigms namely; classification,  NER tagging and question answering. To effectively explore potential of generative language models across all three paradigms,  we have deliberately selected four distinct tasks including requirements extraction \citep{ivanov2022extracting}, requirements classification \citep{saleem2023fnreq},  NER tagging \citep{malik2021named, malik2022software, herwanto2021named, ahmed2021novel} and question answering system \citep{ezzini2023ai, abualhaija2022automated}. Our strategic choice of these tasks is driven by the aim of simplifying the experimental process while ensuring a comprehensive exploration of the capabilities of large language models across various aspects of requirements analysis. Each task presents unique challenges and complexities, collectively providing a well-rounded assessment of the model’s performance and versatility in addressing diverse requirements-related scenarios.}
    \item \added{This study performs a large-scale comparison of ChatGPT and Gemini across four distinct requirement engineering tasks which fall into three working paradigms namely; classification, named entity tagging, and question answering.}
    \item Across 4 different tasks public benchmark datasets, \deleted{it}\added{this study} compares performance of both generative models (ChatGPT, Gemini) with existing machine/deep learning and traditional language models based predictive pipelines, specifically designed for these tasks.
\end{enumerate}
\section{Related work}\label{related-work}
This section explores various predictors designed for 4 distinct requirement engineering applications including: requirements extraction, classification, named entity annotation and domain-specific question-answering system. It provides \added{an} insight into the evolution and advancements of these predictors in their respective application areas.
\subsection{Requirements extraction}
Two different research studies have been conducted to empower the process of requirement extraction from SRS documents by harnessing the benefits of AI \citep{ivanov2021extracting, sainani2020extracting}. Vladimir et al. \citep{ivanov2021extracting} constructed an independent test set by manually annotating the content of SRS documents as requirements and non-requirements. \added{The authors} \deleted{Authors} bench-marked \added{the} performance of independent test set across three different predictors namely: FastText, ELMO and BERT. A thorough experimental analysis revealed that the BERT predictor consistently outperformed \added{the} other two predictors. Sainani et al. \citep{sainani2020extracting} also generated \added{a} requirement extraction dataset using contracts documents. The authors used three traditional machine learning and one deep learning based predictive pipelines namely, support vector machine (SVM), random forest (RF), naive bayes (NB) and Bi-directional LSTM (BiLSTM). Their \citep{sainani2020extracting} experimental analysis demonstrated that BiLSTM surpassed performance of \added{the} other traditional ML based predictive pipelines.

\subsection{Requirements classification}
 Over past 4 years, around 11 different ML and DL based predictors (\citep{althanoon2021supporting, dias2020software, haque2019non, toth2019comparative, rahimi2020ensemble, tiun2020classification, kaur2022sabdm, khayashi2022deep, rahimi2021one, saleem2023fnreq}) have been developed for software requirements classification. Among 11 predictors, 4 ML based predictors \citep{althanoon2021supporting, dias2020software, haque2019non, toth2019comparative}) were developed by utilizing Bag of Words (BoW) based text representation approach with traditional ML classifiers including: logistic regression (LR), SVM, stochastic gradient descent (SGD), decision trees (DT), k-Nearest Neighbors (kNN) and probabilistic NB based models (multinomial, Bernoulli, Gaussian). Furthermore, a meta predictor \citep{rahimi2020ensemble} was developed by reaping the benefit of 5 standalone ML classifiers (LR, SVC, SVM, DT and NB).

Among 5 DL based predictors  \citep{tiun2020classification, kaur2022sabdm, khayashi2022deep, rahimi2021one, saleem2023fnreq}, two predictors \citep{tiun2020classification, kaur2022sabdm} used word embedding approaches namely FastText, Word2Vec and GloVe along with two standalone deep learning architectures namely: long short term memory network (LSTM) and convolutional neural network (CNN). Two predictors \citep{rahimi2021one, khayashi2022deep} made use of random and Glove embeddings for text representation and combined architectures of CNN and LSTM to extract comprehensive contextual features. One predictor \citep{saleem2023fnreq} explored the potential of FastText based text representation along with traditional feature selection and attention mechanism.

Following the success of language model in diverse types of tasks such as biomedical text mining \citep{ji2021dnabert}, text classification \citep{favero2021bert_se} and natural language processing (NLP) tasks \citep{duan2020study}, researchers have investigated the potential of different language models such as BERT \citep{devlin2018bert}, RoBERTa \citep{liu2019roberta}, XLNet \citep{yang2019xlnet}, AL-BERT \citep{lan2019albert} and DistilBERT \citep{sanh2019distilbert} for software requirement classification. In addition to standalone transformer based language models, Li et al., \citep{li2022automatic} proposed an ensemble approach by integrating graph strategy with BERT based language model to reap benefits of both individual approaches. Table  \ref{LS-table-req} presents a comprehensive overview of 11 distinct machine learning and deep learning based requirements classification predictors. \added{Table  \ref{LS-table-req}}\deleted{It} summarises details of datasets used, feature encoding techniques and predictors.
\subsection{Requirements named entity recognition}

Named entity recognition (NER) task annotates words or group of words with predefined tags that refer to particular names, entities, events or actions \citep{mansouri2008named}. In the domain of software engineering, these tags encompass software-specific information including actor, action, operator, user, object, graphical user interface (GUI), hardware, application programming interface (API), etc. NER serves as a backbone for various AI based software engineering pipelines such as information extraction \citep{zhou2020improving}, software requirements classification \citep{dias2020software}, use case diagram generation and topic modeling \citep{kanan2015extracting, arslan2022extracting}. This section briefly describes 3 machine learning, 7 deep learning and 10 language models based NER predictors that have been developed to annotate SRS documents \citep{malik2021named, tabassum2020code, li2019feature}, privacy requirements \citep{herwanto2021named} and software bug specific entity recognition \citep{zhou2020improving, zhou2018recognizing}. 
Among 3 machine learning predictors \citep{vineetha2022multinomial, imam2021svm, vineetha2022passive}, Imam et al. \citep{imam2021svm} predictor utilized linguistic rules and SVM classifier while K V et al. \citep{vineetha2022multinomial} predictor utilized part of speech information along with multinomial naive bayes (MNB) classifier for NER annotations. Moreover, K V et al., \citep{vineetha2022passive} predictor utilized Bag of Words (BoW) approach along with passive-aggressive classifier for NER tagging. 

\begin{table*}[]
\scriptsize
\centering
\caption{A comprehensive summary of existing requirements classification predictors}
\label{LS-table-req}
\renewcommand{\arraystretch}{1.1}
\resizebox{1\textwidth}{!}{
\begin{tabular}{llll}
\hline
\textbf{ \begin{tabular}[c]{@{}l@{}}Author, Year,\\ Ref  \end{tabular}}  & \textbf{Dataset}  & \textbf{\begin{tabular}[c]{@{}l@{}}Feature \\Encoding\end{tabular}} & \textbf{Predictor}   \\ \hline 
\begin{tabular}[c]{@{}l@{}}Saleem et   \\al. 2023,  \citep{saleem2023fnreq} \end{tabular}    
    & \begin{tabular}[c]{@{}l@{}}PROMISE, \\PROMISE-exp\end{tabular}     & FastText & \begin{tabular}[c]{@{}l@{}}Attention based \\ DL Classifier\end{tabular}  \\  \hline  \begin{tabular}[c]{@{}l@{}}Luo et al. \\ 2022,
 \citep{luo2022prcbert} \end{tabular}        & \begin{tabular}[c]{@{}l@{}}PROMISE,\\ NFR-review,\\ NFR-so\end{tabular} & \begin{tabular}[c]{@{}l@{}}BERT\end{tabular}   & 
 \begin{tabular}[c]{@{}l@{}} BERT\end{tabular}  \\ \hline \begin{tabular}[c]{@{}l@{}}Li et al.  \\ 2022, \citep{li2022automatic}\end{tabular}    
  & \begin{tabular}[c]{@{}l@{}}PROMISE, \\ Concordia\end{tabular}     & Node embedding    & \begin{tabular}[c]{@{}l@{}}Graph attention\\ network\end{tabular}     \\ \hline \begin{tabular}[c]{@{}l@{}}Kaur et al.\\ 2022,
 \citep{kaur2022sabdm} \end{tabular}         & \begin{tabular}[c]{@{}l@{}}PROMISE,\\ Open source\\ project\end{tabular}& Glove & \begin{tabular}[c]{@{}l@{}}Self-attention \\ based BiRNN\end{tabular}    \\ \hline \begin{tabular}[c]{@{}l@{}}Ivano et al.\\ 2022,
 \citep{ivanov2022extracting} \end{tabular}     & Pure  & \begin{tabular}[c]{@{}l@{}}FastText, Elmo, \\ BERT\end{tabular}   & SVM, BERT \\ \hline   \begin{tabular}[c]{@{}l@{}}Khayashi et \\ al. 2022,
 \citep{khayashi2022deep}   \end{tabular}     & Pure  & \begin{tabular}[c]{@{}l@{}}Glove, Random \\ word embeddings\end{tabular} & \begin{tabular}[c]{@{}l@{}}CNN, LSTM, \\ BiLSTM, GRU,\\ BiGRU\end{tabular}  \\ \hline
 \begin{tabular}[c]{@{}l@{}}Ajagbe et \\ al. 2022,
 \citep{ajagbe2022retraining} \end{tabular}     & \begin{tabular}[c]{@{}l@{}}PROMISE, \\ Pure,\\ App reviews, \\ Play store\\ reviews\end{tabular} & \begin{tabular}[c]{@{}l@{}}BERT\end{tabular}   & BERT \\ \hline \begin{tabular}[c]{@{}l@{}}kici et al.\\ 2021,
 \citep{kici2021text}  \end{tabular}  & \begin{tabular}[c]{@{}l@{}}Doors,\\PROMISE,\\ Pure \end{tabular}  & \begin{tabular}[c]{@{}l@{}}BERT\end{tabular}   & \begin{tabular}[c]{@{}l@{}}BERT\end{tabular}\\ \hline \begin{tabular}[c]{@{}l@{}}Althanoon et\\ al. 2021,
 \citep{althanoon2021supporting}  \end{tabular}     &PROMISE     &  TFIDF     & MNB, LR   \\ \hline \begin{tabular}[c]{@{}l@{}}Rahimi et \\ al. 2021
 \citep{rahimi2021one}  \end{tabular}       &PROMISE     & \begin{tabular}[c]{@{}l@{}}Random \\ embeddings\end{tabular}  & \begin{tabular}[c]{@{}l@{}}Ensemble of\\ CNN, LSTM, \\ BiLSTM, GRU\end{tabular}   \\ \hline  \begin{tabular}[c]{@{}l@{}}  kici et al.\\ 2021, \citep{kici2021bert}  \end{tabular}   & \begin{tabular}[c]{@{}l@{}}Doors,\\PROMISE\end{tabular}    &  -     & DistillBERT     \\ \hline  \begin{tabular}[c]{@{}l@{}} Favero et al. \\ 2021,
 \citep{favero2021bert_se}   \end{tabular}   & \begin{tabular}[c]{@{}l@{}}Open source\\ project\end{tabular}     & \begin{tabular}[c]{@{}l@{}}BERT\end{tabular}  &  BERT \\ \hline  \begin{tabular}[c]{@{}l@{}} Tiun et al.\\ 2020,
 \citep{tiun2020classification} \end{tabular}   & \begin{tabular}[c]{@{}l@{}}Promise\end{tabular}& Word2vec, FastText   & CNN   \\ \hline   
\begin{tabular}[c]{@{}l@{}}Dias et al. \\ 2020,
 \citep{dias2020software} \end{tabular}  &PROMISE-exp &  TFIDF & \begin{tabular}[c]{@{}l@{}} SVM\end{tabular}    \\ \hline
\begin{tabular}[c]{@{}l@{}} Rahimi et  \\al. 2020,
 \citep{rahimi2020ensemble} \end{tabular}    & Self-collected    & TFIDF& \begin{tabular}[c]{@{}l@{}}Ensemble of SVM, \\ SVC, LR, NB, DT\end{tabular} \\ \hline  \begin{tabular}[c]{@{}l@{}}Hey et al. \\2020,
 \citep{hey2020norbert} \end{tabular}&PROMISE     & \begin{tabular}[c]{@{}l@{}}BERT\end{tabular}  & BERT \\ \hline \begin{tabular}[c]{@{}l@{}} Rehman et al.\\ 2019,
 \citep{rahman2019classifying}   \end{tabular}  &PROMISE     & Word2vec    & RNN, LSTM, GRU  \\ \hline  \begin{tabular}[c]{@{}l@{}} Baker et al.\\ 2019,
 \citep{baker2019automatic} \end{tabular}   &PROMISE     & \begin{tabular}[c]{@{}l@{}}Random \\ embeddings\end{tabular} & ANN, CNN  \\ \hline \begin{tabular}[c]{@{}l@{}} Haque et al.\\ 2019,

 \citep{haque2019non} \end{tabular} &PROMISE     &  TFIDF & NB, SVM, DT, KNN\\ \hline 
\begin{tabular}[c]{@{}l@{}}Toth et al. \\2019,
 \citep{toth2019comparative} \end{tabular} & \begin{tabular}[c]{@{}l@{}}Promise, \\ Stackoverflow\end{tabular}  & TFIDF& \begin{tabular}[c]{@{}l@{}}BernoulliNB, DT\\ ET, ETs, KNN,\\ Label propagation,\\ Label spread, LR, \\ MLP, MNB, SVM\end{tabular} \\ \hline  
\end{tabular}}
\end{table*}

Among 7 deep learning  predictors \citep{kocerka2022ontology, herwanto2021named, malik2021named, pudlitz2019extraction, zhou2020improving, li2019feature, zhou2018recognizing}, Li et al. \citep{li2019feature} predictor utilized Word2Vec word embedding approach along with Bi-LSTM for software requirement NER tagging. Zhou et al. \citep{zhou2018recognizing} also made use of Word2Vec word embedding approach and Bi-LSTM along with linear chain Conditional Random Fields approach (BiLSTM-CRF) for software bugs-specific NER tagging. \deleted{Afterward,} Zhou et al. \citep{zhou2020improving} extended \deleted{their work}\added{the work in \citep{zhou2018recognizing}} by utilizing Word2Vec word embedding approach and deep BiLSTM-CRF model along with attention mechanism for bugs specific NER annotation. On top of Word2Vec word embedding approach, Pudlitz et al. \citep{pudlitz2019extraction} designed a hybrid predictor  by utilizing two different neural architectures, namely convolutional neural networks (CNNs) and bidirectional long short term memory networks (Bi-LSTMs). Moreover, Malik et al. \citep{malik2021named} end to end pipeline comprises GloVe word embedding and BiLSTM-CRF while Herwanto et al. \citep{herwanto2021named} predictor reaps combine benefits of two different word embedding approaches (GloVe and Flair embedding) and BiLSTM-CRF.  Kocerka et al., \citep{kocerka2022ontology} explored the potential of FastText word embedding and convolutional neural network (CNN).

Following success of language models in diverse types of NLP tasks \citep{min2023recent} \citep{gu2021domain}, Ray et al. \citep{tikayat2023aerobert} fine-tuned English Wikipedia and BookCorpus based pre-trained BERT language model on aerospace domain related requirements. Moreover, Chow et al. \citep{chow2023analysis} made use of BookCorpus and English Wikipedia content based pre-trained BERT language model for embedded systems domain related requirements while Lopez et al. \citep{lopez2021mining} explored potential of two language models ($BERT_{base}\-CRF$, $SciBERT\-CRF$) for software requirement NER annotation. Tabassum et al. \citep{tabassum2020code} utilized Stack Overflow and GitHub corpus contents to train BERT language model in an unsupervised fashion. They explored the potential of domain specific pre-trained model for programming domain related requirements NER tagging. Moreover, Malik et al. \citep{malik2023transfer} predictor explored the potential of Sentence-BERT (SBERT) which is a variant of pre-trained BERT language model and utilizes Siamese and triplet network structures to generate semantically meaningful sentence embeddings. They fine-tuned SBERT for software, hardware, healthcare and transportation domains related requirement NER annotation. Zhou et al., \citep{zhou2022named} proposed BERT-GloP-Rule predictor that reaps the benefits of both BERT and global pointers for NER annotations. Afterward, Malik et al. \citep{malik2022identifying} utilized pre-trained BERT and TFIDF representations for development of NER tagger. Afterward, Malik et al., \citep{malik2022software} employed pre-trained BERT language model and fine-tuned it on three requirement datasets (DOORS, SRE, and RQA) for NER annotation. Das et al., \citep{das2023zero} explored the potential of zero-shot learning strategy along with T5 language model. Moreover, Tang et al., \citep{tang2023attensy} predictor utilized BERT pre-trained language model representations along with three different architectures, namely BiLSTM, Graph Convolutional Network (GCN) and attention mechanism. Table \ref{ls-table-ner} presents a comprehensive overview of 20 distinct NER predictors. It summarises details of used datasets, feature encoding techniques and NER predictors.

\begin{table*}[]
\caption{A high-level overview of existing requirements named entity recognition predictors in terms of used datasets, representation learning strategies and predictors.}
\label{ls-table-ner}
\centering
\renewcommand{\arraystretch}{1.4}
\resizebox{\textwidth}{!}{
\begin{tabular}{|c|c|c|c|}
\hline
\textbf{Author, Year {[}ref{]}} & \textbf{Dataset} & \textbf{Feature Encoding Technique} & \textbf{Predictor} \\
\hline
Ray et al., 2023 \citep{tikayat2023aerobert} & Aerospace corpus & BERT & BERT \\
\hline
Das et al., 2023 \citep{das2023zero} &PROMISE & - & T5 \\
\hline
Malik et al., 2023 \citep{malik2023transfer} & - & - & BERT+ Biencoder \\
\hline
Tang et al., 2023 \citep{tang2023attensy} & Stack Overflow & BERT & BiLSTM, GCN \\
\hline
Chow et al., 2022 \citep{chow2023analysis} & GLUE  & - & BERT \\
\hline
Malik et al., 2022 \citep{malik2022software} & DOORS, SRE, RQA  & - & BERT \\
\hline
Zhou et al., 2022 \citep{zhou2022named} & Military requirements & - & BERT + global pointers \\
\hline
Malik et al., 2022 \citep{malik2022identifying} & OpenCoss, WordVista, UAV,  SRS & BERT+TFIDF & BERT \\
\hline
Kocerka et al., 2022 \citep{kocerka2022ontology} & - & FastText & CNN \\
\hline
K V et al., 2022 \citep{vineetha2022multinomial} & NER tags & - & MNB \\
\hline
Imam et al., 2022 \citep{imam2021svm} & CoNLL-2009 & Linguistic rules & SVM \\
\hline
K V et al., 2022 \citep{vineetha2022passive} & NER tags & TFIDF & Passive Aggressive classifier \\
\hline
Herwanto et al., 2021 \citep{herwanto2021named} & Dalpiaz user stories  & GloVe \& Flair embedding & BiLSTM-CRF \\
\hline
Malik et al., 2021 \citep{malik2021named} & DOORS & GloVe & BiLSTM-CRF \\
\hline
Lopez et al., 2021 \citep{lopez2021mining} & SoftCite & - & BERT-CRF \\
\hline
Tabassum et al., 2020 \citep{tabassum2020code} & Stack Overflow NER & - & BERT \\
\hline
Zhou et al., 2020 \citep{zhou2020improving} & Mozilla, Eclipse, Apache, Kernel & Word2Vec & BiLSTM-CRF \\
\hline
Li et al., 2019 \citep{li2019feature} & Chinese Stack Overflow & Word2Vec & BiLSTM \\
\hline
Pudlitz et al., 2019 \citep{pudlitz2019extraction} & German automotive requirements & Word2Vec & CNN, BiLSTM \\
\hline
Zhou et al., 2018 \citep{zhou2018recognizing} & Mozilla, Eclipse & Word2Vec & BiLSTM \\
\hline
\end{tabular}}
\end{table*}

\subsection{Requirements question answering system}
A domain-specific question answering system (QAS) provides precise information within a particular subject area or field. Software systems domain-specific QAS provides useful insights about domain specific knowledge such as paradigm of software development life  \citep{smart2023bdd, pothukuchi2023impact}, requirement analysis \citep{sangaroonsilp2023taxonomy,abualhaija2023legal} and test case generation \citep{hashtroudi2023automated, lukasczyk2023empirical}. This system facilitates researchers and software developers by providing precise and relevant answers of their questions \citep{lende2016question, mervin2013overview}.

This section describes 4 different transformers based predictors \citep{ezzini2023ai, abualhaija2022automated, abualhaija2022coreqqa, do2021towards} for software requirements related QA task. Among 4 different studies, 2 studies \citep{abualhaija2022automated, abualhaija2022coreqqa} utilized GDPR dataset that contains questions and answers related to software privacy requirements. Among remaining two studies\added{,} one study \citep{ezzini2023ai} utilized QAssist dataset that contains requirements related to security, aerospace and defence\added{,} while other study \citep{do2021towards} used stack-overflow based documents to answer development question of Java language.

Among existing predictors, one predictor \citep{ezzini2023ai} explored the potential of 5 different language models (BERT \citep{devlin2018bert}, ALBERT \citep{lan2019albert}, Distil-BERT \citep{sanh2019distilbert}, RoBERTa \citep{liu2019roberta}, ELECTRA \citep{clark2020electra}) for finding relevant answer to questions related to software requirements. Afterwards, Abualhaija et al., \citep{abualhaija2022automated} explored combined potential of 4 language models (BERT \citep{devlin2018bert}, Al-BERT \citep{lan2019albert}, RoBERTa \citep{liu2019roberta}, ELECTRA \citep{clark2020electra}) along with two similarity measures (cosine and BERT cross-encoder similarity) for QA related to software requirements. Abualhaija et al., \citep{abualhaija2022coreqqa} developed COREQQA predictor that explored combined potential of BERT cross-encoder (BCE) and RoBERTa for question answering task. 
\newline
\added{A comprehensive analyses of existing literature reveals that researchers have explored potential of diverse types of machine, deep learning, and foundational language models for distinct types of requirement engineering tasks. However, the potential of advance generative language models is not fully explored. Primary, objective of this study is to explore generative language models potential across distinct types of requirement engineering tasks.}

\section{Materials and methods}
This section summarizes \deleted{specific}\added{the} criteria\deleted{,} utilized to \deleted{deeply} explore the potential of generative language models across four different requirement engineering tasks. Furthermore, it provides high level overview of benchmark datasets and evaluation measures\deleted{,} used to evaluate performance of ChatGPT and Gemini.
\begin{figure*}[]
 \centering 
 \includegraphics[width=1\textwidth]{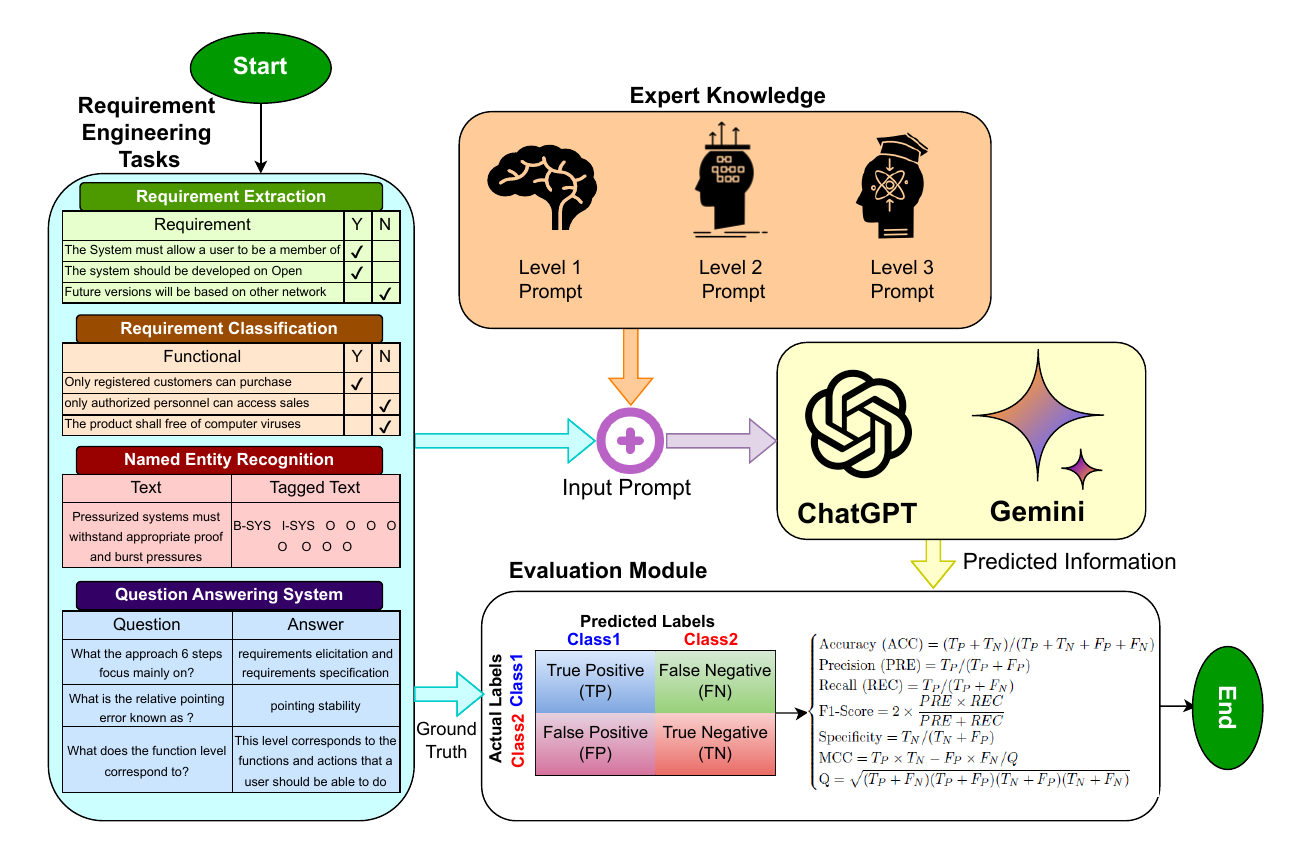}
 \caption{Graphical illustration of prompt engineering for generative language models utilization and evaluation for four distinct tasks including requirements extraction, classification, name entity recognition and question answering system}
\label{methodology}
\end{figure*}

\subsection{Proposed methodology}
Figure \ref{methodology} graphically represents utilization and evaluation of generative language models across four different tasks including: requirements extraction, classification, named entity recognition and question answering system. \deleted{It can be seen in Figure \ref{methodology}, four different tasks related input along with} \added{Figure \ref{methodology}, graphically illustrates input of four different tasks and} three different levels \added{of} expert knowledge is fed to generative language models. Expert level 1 knowledge prompt seeks to assess, how precisely generative language models fulfils desired need by utilizing its background knowledge. At this level, the model is fed with simple query and its performance is evaluated based on the output. Expert level 2 knowledge prompt aims to provide comprehensive understanding to generative language models and output is evaluated based on provided information and background knowledge. Similarly, expert level 3 knowledge prompt provides in-depth task knowledge with relevant example to evaluate \added{the} output of generative language models. Tables 1, 2, 3 and 4 of supplementary file present three different expert level knowledge prompts for requirements extraction, classification, NER tagging and question-answering systems, respectively.

\subsection{Benchmark datasets}\label{dataset-section}
This section summarizes detail of 4 benchmark datasets that are utilized to evaluate the performance of generative language models.

\subsubsection{Requirements extraction}
To facilitate development of AI applications for requirement extraction task, Vladimir et al. \citep{ivanov2021extracting} developed a benchmark dataset by acquiring content from 79 different SRS documents. Authors manually annotated SRS documents content into requirements and non requirements classes. Moreover, the authors collaborated with domain expert to resolve ambiguous cases and developed 4,145 requirement class samples and 3,600 non-requirements class samples.

\subsubsection{Requirements classification}
To evaluate the performance of GLMs for requirements classification we utilized \deleted{promise}\added{PROMISE} benchmark dataset \citep{sayyad2005promise}. PROMISE dataset comprises of 625 requirements including; 254 functional and 371  non-functional requirements.

\subsubsection{Requirements named entity recognition}
To empower aerospace requirements engineering computer-aided programs, Tikayat et al. \citep{tikayat2023aerobert}  prepared a NER dataset containing two distinct types of content: generic aerospace content and aerospace-based requirements.  \deleted{The former}\added{Generic aerospace content} includes scholarly articles and papers published by the National Academy of Space Studies Board \footnote{https://www.nationalacademies.org/ssb/space-studies-board}. These documents provide high-level overview of aerospace domain. \deleted{The latter}\added{On the other hand, aerospace-based requirements} encompass information about standards and requirements for certifying new aircraft designs. Overall, the dataset contains 1432 sentences, including 1,100 sentences about general information, definitions and design methodologies of aerospace domain. The remaining 332 sentences contain aerospace requirements covering diverse range of topics such as environmental factors, performance criteria, functionality, quality assurance, design process and interface.

Furthermore, to optimize NER analysis, authors carefully reformatted these sentences. They eliminated non-textual elements (figures, tables and equations) from original sources to  solely focus on textual data. In order to prevent misinterpretation of sentence boundaries, periods that did not signify end of sentence were replaced with dashes. Additionally, `§' symbol, often used to denote sections in legal and technical documents was replaced with word `Section'.  The authors used `BIO' tagging scheme to label the dataset, where  `B' denotes beginning of an entity, `I'  denotes inside an entity and `O' denotes outside an entity. The authors tagged entities into five main categories, which are selected based on their occurrence frequencies in aerospace corpus. These categories include: Systems (SYS), Values (VAL), Dates and Times (DATETIME), Organizations (ORG) and Resources (RES).  The details of aforementioned tags are provided in Table \ref{NER-tag}.

\begin{table*}[]
\centering
\caption{NER tags and their count in the dataset}
\label{NER-tag}
\renewcommand{\arraystretch}{1.8}
\resizebox{1\textwidth}{!}{
\begin{tabular}{|l|c|c|c|c|c|c|c|c|c|c|c|}
\hline
\textbf{Label} & 0     & 1 & 2& 3     & 4     & 5   & 6   & 7 & 8  & 9   & 10 \\ \hline
\textbf{NER Tag}     & O     & B-SYS   & I-SYS  & B-VAL & I-VAL & B-DATETIME& I-DATETIME& B-ORG   & I-ORG    & B-RES     & I-RES    \\ \hline
\textbf{Description} & Other & \multicolumn{1}{l|}{\begin{tabular}[c]{@{}l@{}}Beginning \\ of a System\end{tabular}} & \multicolumn{1}{l|}{\begin{tabular}[c]{@{}l@{}}Inside\\ a System\end{tabular}} & \multicolumn{1}{l|}{\begin{tabular}[c]{@{}l@{}}Beginning\\ of a Value\end{tabular}} & \multicolumn{1}{l|}{\begin{tabular}[c]{@{}l@{}}Inside\\ a Value\end{tabular}} & \multicolumn{1}{l|}{\begin{tabular}[c]{@{}l@{}}Beginning\\ of a Date-Time\end{tabular}} & \multicolumn{1}{l|}{\begin{tabular}[c]{@{}l@{}}Inside a\\ Date-Time\end{tabular}} & \multicolumn{1}{l|}{\begin{tabular}[c]{@{}l@{}}Beginning\\ of an Organization\end{tabular}} & \multicolumn{1}{l|}{\begin{tabular}[c]{@{}l@{}}Inside an \\ Organization\end{tabular}} & \multicolumn{1}{l|}{\begin{tabular}[c]{@{}l@{}}Beginning \\ of a Resource\end{tabular}} & \multicolumn{1}{l|}{\begin{tabular}[c]{@{}l@{}}Inside a\\ Resource\end{tabular}} \\ \hline
\textbf{Count} & 37686    & 1915 & 1104& 659    & 507    & 147  & 63   & 302 & 227  & 390  & 1033\\ \hline
\end{tabular}
}
\end{table*}

\subsubsection{Requirments based question answering}
 Ezzini et al. \citep{ezzini2023ai} developed a public benchmark software requirement related question answering specific dataset (REQuestA), by utilizing Wikipedia articles and data from six SRSs related to three different domains, namely: (a) aerospace, (b) defence and (c) security. The authors automatically generated question answering pairs using text generation models and then verified data by two domain experts.

The process of generating question answer pairs can be described in 4 different steps. First step involves pre-processing of SRSs by identifying keywords for analyzing domain of SRSs. The authors utilized REGICE \citep{arora2016automated} tool for extracting software specific keywords.  Afterwards, they grouped SRS documents into 3 distinct domains based on TFIDF score of distinct concepts. In order to compute TFIDF score, they preferred to work over phrases rather than individual words. Afterwards, they sorted computed score in descending order and selected top 50 concepts or keywords. Finally, they used each keyword to find a matching Wikipedia article and randomly selected matching articles. Third steps involves automatic splitting of SRSs and Wikipedia articles into text passages. Forth step used a question generation (QG) model (i.e. T5 language model). QG model takes text passage as input and extracts random answer for each passage. Based on generated answer, it automatically generates corresponding question. The output of each text passage involves that passage along with a set of automatically generated question answers pairs \deleted{<q,a>}. In order to validate question answer pairs, authors used two methods: (1) automatic validation by question answer evaluator and (2) manual validation by domain experts. In first validation process, author utilized BERT based question answer evaluator which takes question answer pair as input and output validity score of each pair. These QA pairs are then sorted based on validity score and top 5\% of pairs are selected. Second validation method involves verification from 2 third party domain experts. The list of automatically generated QA pairs, original SRSs documents and randomly selected Wikipedia articles are shared with both experts; who manually validated QA pairs associated with each text passage. 

REQuestA dataset contains 387 QA $<q, a>$ pairs. Among these QA pairs, 103 pairs from SRSs and 111 pairs from Wikipedia (214 total question answer pairs) were manually defined by experts while 86 pairs from SRSs and 87 pairs from Wikipedia articles were automatically generated using QA  automatic evaluators.

\subsection{Evaluation measures}
Following evaluation criteria of existing requirement extraction, requirement classification and NER tagging studies, generative language models performance is computed using 4 different evaluation measures including: accuracy\citep{hassan2019story, summra2021supervised}, precision, recall and f1-score. These measures compute \added{the} predictors performance by utilizing confusion matrix shown in Figure \ref{methodology} evaluation module. The confusion matrix contains 4 different types information namely, true positive ($T_{pos}$), true negative ($T_{neg}$), false positive ($F_{pos}$) and false negative ($F_{neg}$) to calculate overall performance of predictor. True positive ($T_{pos}$) represents number of positive class instances  that are correctly identified by predictor. Similarly, true negative ($T_{neg}$) denotes number of negative class instances that are correctly identified by predictor. On the other hand, false positive ($F_{pos}$) reflects number of negative class instance that are incorrectly identified by predictor as positive class. False negative ($F_{neg}$) denotes number of negative class instance that are incorrectly identified by predictor as positive class. Furthermore, mathematical expressions of 4 different evaluation measures are also shown in Figure \ref{methodology} evaluation module.  

Furthermore, adhering to evaluation criteria of existing question answer systems  studies \citep{chen2019evaluating, risch2021semantic, akermi2020transformer, butt2021transformer}\added{,} we use 8 different evaluation measures namely\deleted{;}\added{:} precision, recall, f1-score, ROUGE-1, Rouge-2, Rouge-L, ROUGE-S, METEOR (Metric for Evaluation of Translation with Explicit ORdering) score \citep{lavie2009meteor} and BLEU (BiLingual Evaluation Understudy) score \citep{papineni2002bleu}\deleted{score}.  ROUGE-1 calculates the ratio  of overlapping unigrams in  model-generated output and ground truth divided by total number of unigrams in model-generated output. Similarly Rouge-2 computes the ratio of matching bigrams in model-generated output that also appear in ground truth, over total number of bigrams in model-generated output. Rouge-L computes longest common sub-sequence that is shared between model generated output and ground truth. A longer shared sequence should indicate more similarity between the two sentences. ROUGE-S is a skip-gram concurrence metric, that allows to  identify consecutive words from ground truth that also appear in model-generated output despite being separated by one or more other words. BLEU score  is computed by comparing model-generated output with ground truth text. It involves tokenization, counting matching n-grams, calculating precision for different n-gram orders, introducing a brevity penalty, and combining precisions using a weighted geometric mean. The formula for BLEU score is product of the brevity penalty and  exponential of average log precision. METEOR score is computed by aligning model-generated output with ground truth, using various linguistic resources and matching criteria. Unlike BLEU score it not only considers unigrams precision but also considers additional linguistic feature like recall stemming and synonym. It involves  calculation of precision, recall and an alignment penalty, which penalizes non-matching words. The final METEOR score is a harmonic mean of precision and recall, adjusted by the alignment penalty. Equation 1 of supplementary file illustrates mathematical formulation to compute aforementioned measures.

\section{Experimental setup and results}
We developed python language scripts in order to fed both GLMs APIs: (ChatGPT (3.5) \footnote{https://chat.openai.com/} and Gemini \footnote{https://www.gemini.com/}) with three different levels expert knowledge based prompts. In response to input prompts both GLMs produced diverse types of content for all four tasks. For classification tasks, we directly extracted desired label from generated output. However, output of GLMs for NER task posed significant challenges to extract original content and predicted tags. To automate evaluation of NER task, we closely analyzed output of both GLMs and wrote a rule based content and tags extractor script. Furthermore, for question answering system, we took all generated content to assess GLMs performance.

To facilitate fair performance analysis, we adopted evaluation criteria of existing studies \citep{saleem2023fnreq,  ivanov2021extracting, tikayat2023aerobert, ezzini2023ai} to compare effectiveness of both GLMs with existing predictors. For requirement extraction and classification benchmark datasets \citep{saleem2023fnreq,  ivanov2021extracting, tikayat2023aerobert, ezzini2023ai}, authors have provided standard train\textbackslash test splits and we utilized only test sets to compute GLM's performance. \deleted{However, for} \added{For}  NER benchmark dataset, authors of existing study \citep{tikayat2023aerobert} utilized 90\% data for training and 10\% data for evaluation. However, authors \citep{tikayat2023aerobert} did not provide separate train \added{and} test sets. To mitigate biased evaluation\deleted{,} we performed five random data splits and averaged their performance for comparison with existing predictor. In question answering system \citep{ezzini2023ai}, similar to evaluation criteria of \added{the} existing study, we utilized all questions to \added{the} acquired answers from both GLMs.

\subsection{Results}
This section comprehensively examines impact of prompt engineering on performance of two GLMs across four distinct requirement engineering tasks. Additionally, it provides thorough  performance analysis of GLMs with exiting machine learning, deep learning and language model based predictors of four tasks across benchmark datasets.

\subsubsection{Generative language models performance analysis with multiple prompts}

\begin{table}[]
\caption{A visual analysis of confusion matrices and performance metrics of generative language models for requirements extraction task.}
\label{RE-CM}
\renewcommand{\arraystretch}{1.5}
\resizebox{\textwidth}{!}{
\begin{tabular}{|l|c|c|c|c|c|c|c|c|c|c|}
\hline
\multirow{2}{*}{\begin{tabular}[c]{@{}l@{}}Prompts with Different \\ Expert Knowledge\end{tabular}} & \multicolumn{5}{|c|}{Gemini}    & \multicolumn{5}{|c|}{ChatGPT}  \\ \XGap{-2.5pt} \cline{2-11} \XGap{-5.5pt}
    & Accuracy & Precision & Recall & F1-Score & MCC   & Accuracy & Precision & Recall & F1-Score & MCC   \\ \hline
Prompt 1               & 0.746    & 0.758     & 0.746  & 0.750    & 0.431 & 0.77     & 0.761     & 0.77   & 0.762    & 0.437 \\ \hline
Prompt 2               & 0.753    & 0.761     & 0.753  & 0.756    & 0.441 & 0.743    & 0.737     & 0.743  & 0.739    & 0.385 \\  \hline
Prompt 3               & 0.790    & 0.785     & 0.790  & 0.774    & 0.475 & 0.75     & 0.745     & 0.75   & 0.747    & 0.403 \\ \hline
  & \multicolumn{5}{c|}{}    & \multicolumn{5}{c|}{} \\  
 \begin{tabular}[c]{@{}l@{}}   Confusion Matrix \\ \hspace{0.2in} \\ \hspace{0.2in} \\ \hspace{0.2in}  \end{tabular}  & \multicolumn{5}{c|}{\includegraphics[width=0.50\textwidth]{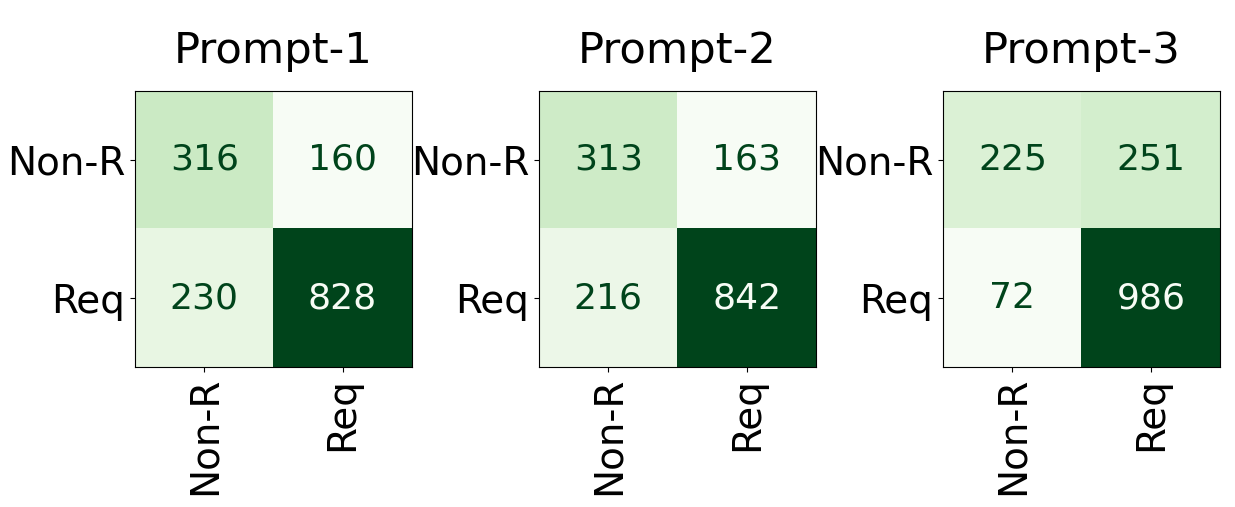}}    & \multicolumn{5}{c|}{\includegraphics[width=0.50\textwidth]{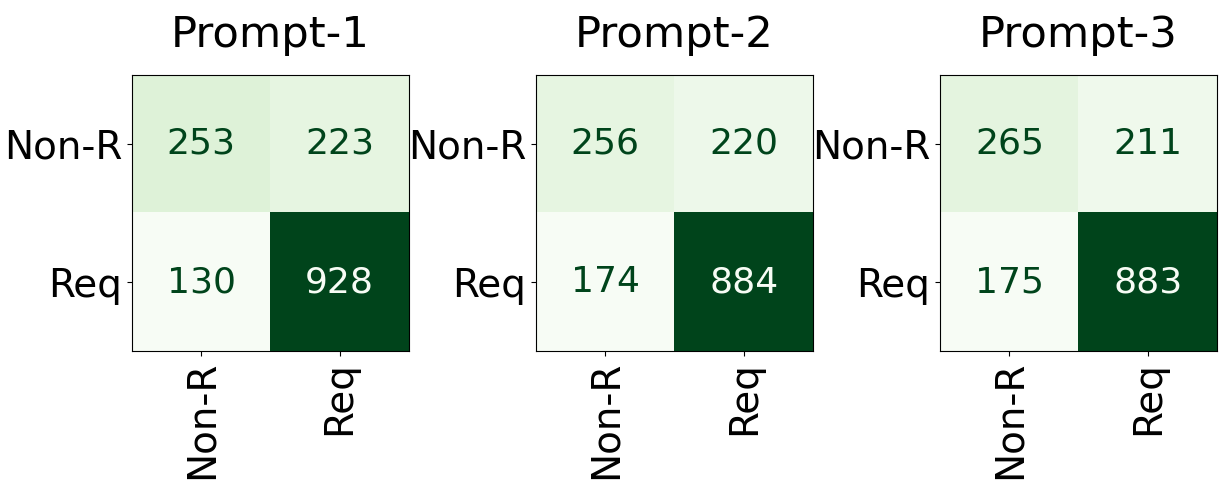}} \\ \hline         
\end{tabular}}

\end{table}
This section investigates impact of prompt engineering on performance of  GLMs(ChatGPT, Gemini) across four distinct requirement engineering tasks. In three tasks including requirements extraction, requirements classification and named entity recognition prompts are designed by incorporating three different level expert knowledge about the task\deleted{,}\added{.} \deleted{such as}\added{At} level 1 expert knowledge (low knowledge)\deleted{,} prompts contains basic information about underlying task. At level 2 expert knowledge (medium knowledge) \added{the} prompts contain background knowledge about tasks with comprehensive definitions\added{.} \deleted{w}\added{W}hile at level 3 expert knowledge (expert knowledge) the prompts encompass tasks background knowledge with definitions and examples. In \added{the} question answering task, \added{the} first prompt only contain questions and \deleted{the} other two prompts along with questions contain text that contain answers of questions. 

Table \ref{RE-CM} presents \added{the} performance metrics and confusion matrices achieved by GLMs(ChatGPT, Gemini) when they are fed with three distinct prompts, across requirement extraction benchmark dataset. Gemini produced accuracy values of 0.746, 0.753 and 0.79 at prompt 1, prompt 2, and prompt 3 respectively. These values reveals Gemini's performance consistently improved as the prompts are enriched with higher levels of expert knowledge.  In contrast, ChatGPT's performance exhibited variations with increasing prompt expertness. While it achieved the highest accuracy of 0.77 at prompt 1, accuracy values of 0.743 and 0.75 are produced at prompts 2 and 3, respectively. Furthermore, confusion metric analysis of Table \ref{RE-CM}, reveals that with increased expert knowledge, Gemini becomes biased towards requirements class while ChatGPT demonstrates a biasness towards non-requirements class.

\begin{table*}[]
\caption{A visual analysis of confusion matrices and performance metrics of generative language models for requirements classification task.}
\label{RC-CM}
\renewcommand{\arraystretch}{1.5}
\resizebox{1\textwidth}{!}{
\begin{tabular}{|l|c|c|c|c|c|c|c|c|c|c|}
\hline
\multirow{2}{*}{\begin{tabular}[c]{@{}l@{}}Prompts with Different \\ Expert Knowledge\end{tabular}} & \multicolumn{5}{c|}{Gemini}     & \multicolumn{5}{c|}{ChatGPT}    \\ \XGap{-2.5pt} \cline{2-11} \XGap{-5.5pt}
   & Accuracy & Precision & Recall & F1-Score & MCC   & Accuracy & Precision & Recall & F1-Score & MCC   \\ \hline
Prompt 1 & 0.664    & 0.740     & 0.664  & 0.661    & 0.408 & 0.744    & 0.821     & 0.744  & 0.743    & 0.566 \\ \hline
Prompt 2 & 0.784    & 0.796     & 0.784  & 0.786    & 0.569 & 0.784    & 0.817     & 0.784  & 0.786    & 0.596 \\ \hline
Prompt 3 & 0.784    & 0.796     & 0.784  & 0.786    & 0.569 & 0.712    & 0.806     & 0.712  & 0.708    & 0.523 \\ \hline 
  & \multicolumn{5}{c|}{}    & \multicolumn{5}{c|}{} \\  
 \begin{tabular}[c]{@{}l@{}}   Confusion Matrix \\ \hspace{0.2in} \\ \hspace{0.2in} \\ \hspace{0.2in}  \end{tabular}  & \multicolumn{5}{c|}{\includegraphics[width=0.50\textwidth]{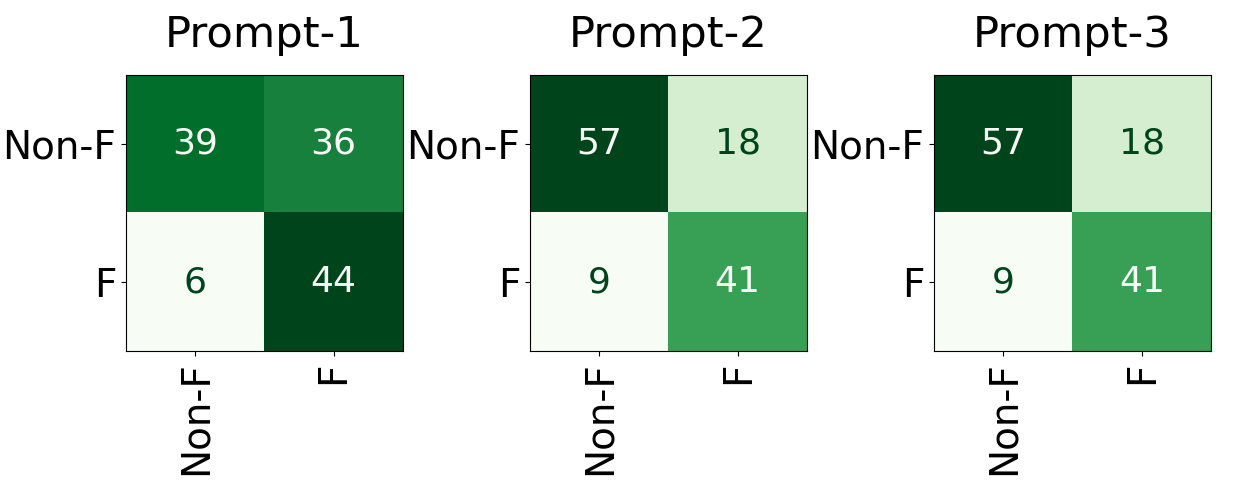}}    & \multicolumn{5}{c|}{\includegraphics[width=0.50\textwidth]{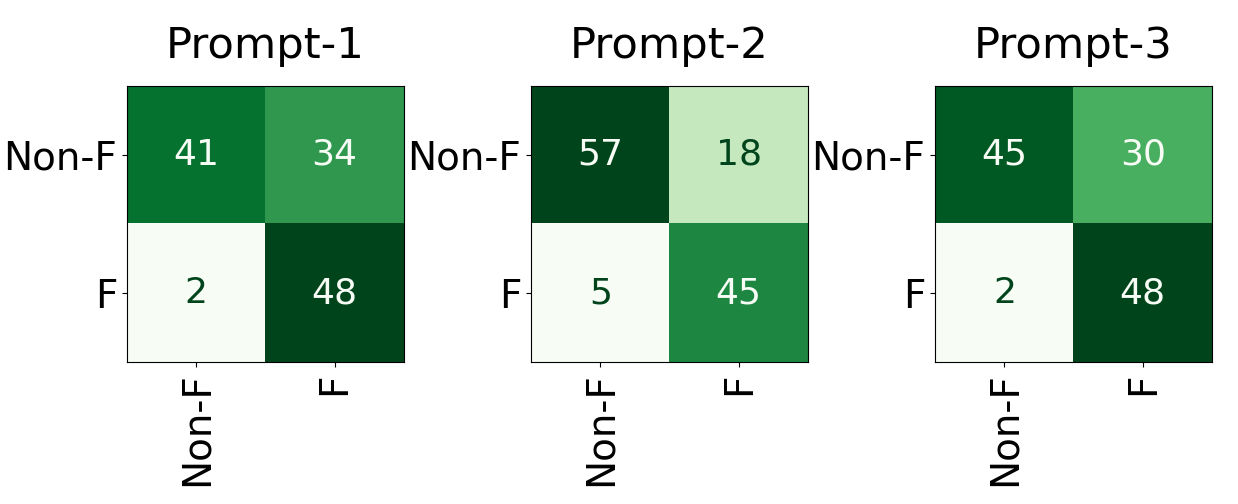}} \\ \hline  
\end{tabular}}
\end{table*}

Table \ref{RC-CM} illustrates performance comparison and confusion matrices of ChatGPT and Gemini based on 3 distinct prompts across benchmark dataset of requirements classification task.  Gemini secures accuracy of 0.66 at prompt 1 while achieves same accuracy score of 0.78 at prompts 2 and 3. Contrarily, ChatGPT  shows inconsistent performance with varying prompts and achieves lowest accuracy of 0.71 at prompt 3 and highest accuracy of 0.78 at prompt 2. The confusion matrices demonstrates that with increased expert knowledge Gemini is capable of correctly identifying non-requirement. Contrarily, ChatGPT exhibit biasness towards requirement class at prompts 1 and 3 and wrongly identifies non-requirement samples as requirements while at prompt 2 it managed to correctly identify non-requirements samples.



Table \ref{NER-CM} illustrates GLMs's performance across 3 different prompts, in terms of 3 different evaluation measures and confusion matrices for requirements NER tagging task benchmark dataset. It can be seen in Table  \ref{NER-CM}, that ChatGPT consistently beats Gemini  to accurately identifying temporal, value and organizational elements (DATETIME , VAL and ORG). Moreover, both GLMS consistently miss-classify SYS as RES  and RES is  frequently confused with other classes (SYS, ORG and Val). Furthermore, the number of NER tags miss-classified as OTHER increases with incorporation of task specific information in prompts. Hence, context rich information introduces complexities resulting in higher likelihood of miss-classifying named entities as ``OTHER". Overall, Gemini shows best performance with prompt 1 while ChatGPT exhibits peak performance using prompt 2.  Hence, incorporation of domain specific knowledge deteriorate Gemini's performance but enhances ChatGPT performance at a certain level (i.e medium level expert knowledge).
          
\begin{table*}[]
\caption{A visual analysis of confusion matrices and performance metrics of generative language models for named entity recognition of requirements.}
\label{NER-CM}
\renewcommand{\arraystretch}{1.5}
\resizebox{1.0\textwidth}{!}{
\begin{tabular}{|l|l|cccc|cccc|}
\hline
\multirow{2}{*}{\textbf{Prompts}} & \multicolumn{1}{c|}{\multirow{2}{*}{\begin{tabular}[c]{@{}c@{}}\textbf{NER} \\ \textbf{Tags}\end{tabular}}} & \multicolumn{4}{c|}{\textbf{Gemini}}           & \multicolumn{4}{c|}{\textbf{ChatGPT}}           \\ \XGap{-2.5pt} \cline{3-10} \XGap{-5.5pt}  
&   & \multicolumn{1}{c|}{\textbf{Precision}} & \multicolumn{1}{c|}{\textbf{Recall}} & \multicolumn{1}{c|}{\textbf{F1-Score}}    & \textbf{Confusion Matrix}         & \multicolumn{1}{c|}{\textbf{Precision}} & \multicolumn{1}{c|}{\textbf{Recall}} & \multicolumn{1}{c|}{\textbf{F1-Score}}    & \textbf{Confusion Matrix}            \\ \hline
\multicolumn{1}{|c|}{\multirow{6}{*}{ \rotatebox[origin=c]{90}{Prompt 1}}} & DATETIME  & \multicolumn{1}{c|}{43.21}     & \multicolumn{1}{c|}{46.58}  & \multicolumn{1}{c|}{43.66}  & \multirow{6}{*}{\includegraphics[width=0.20\textwidth]{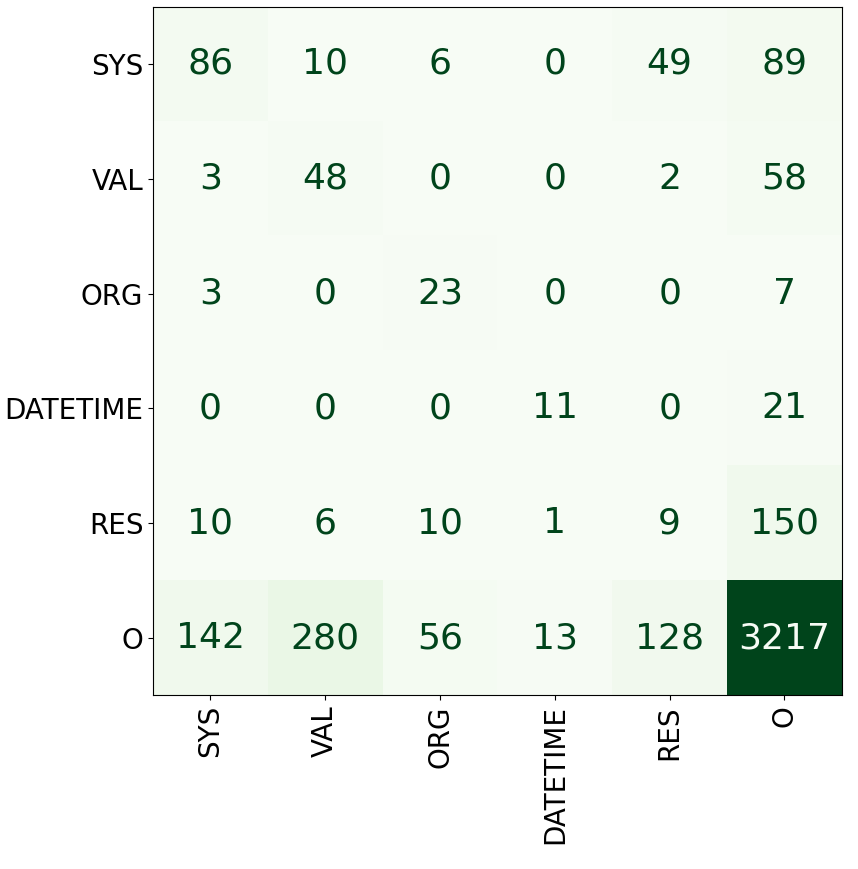}} & \multicolumn{1}{c|}{43.23}     & \multicolumn{1}{c|}{61.17}  & \multicolumn{1}{c|}{50.174} & \multirow{6}{*}{\includegraphics[width=0.20\textwidth]{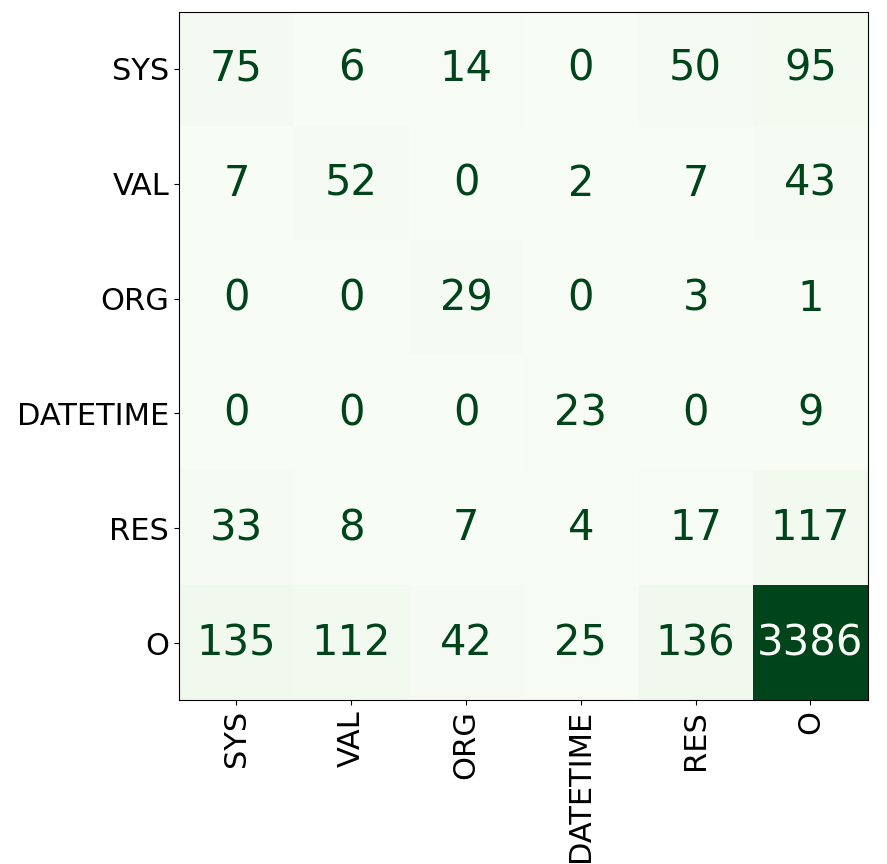}} \\ \XGap{-2.5pt} \cline{2-5} \cline{7-9} \XGap{-5.5pt}
\multicolumn{1}{|c|}{}                          & ORG       & \multicolumn{1}{c|}{27.6}      & \multicolumn{1}{c|}{74.72}  & \multicolumn{1}{c|}{40.19}  &                          & \multicolumn{1}{c|}{32.34}     & \multicolumn{1}{c|}{72.31}  & \multicolumn{1}{c|}{44.56}  &                             \\ \XGap{-2.5pt} \cline{2-5} \cline{7-9} \XGap{-5.5pt}
\multicolumn{1}{|c|}{}                          & RES       & \multicolumn{1}{c|}{1.19}      & \multicolumn{1}{c|}{3.91}   & \multicolumn{1}{c|}{1.82}   &                          & \multicolumn{1}{c|}{2.14}      & \multicolumn{1}{c|}{6.18}   & \multicolumn{1}{c|}{3.15}   &                             \\ \XGap{-2.5pt} \cline{2-5} \cline{7-9} \XGap{-5.5pt}
\multicolumn{1}{|c|}{}                          & SYS       & \multicolumn{1}{c|}{0.299}     & \multicolumn{1}{c|}{28.02}  & \multicolumn{1}{c|}{28.92}  &                          & \multicolumn{1}{c|}{0.26398}   & \multicolumn{1}{c|}{21.31}  & \multicolumn{1}{c|}{23.5}   &                             \\ \XGap{-2.5pt} \cline{2-5} \cline{7-9} \XGap{-5.5pt}
\multicolumn{1}{|c|}{}                          & VAL       & \multicolumn{1}{c|}{8.29}      & \multicolumn{1}{c|}{26.06}  & \multicolumn{1}{c|}{12.54}  &                          & \multicolumn{1}{c|}{19.97}     & \multicolumn{1}{c|}{29.2}   & \multicolumn{1}{c|}{23.712} &                             \\ \XGap{-2.5pt} \cline{2-5} \cline{7-9} \XGap{-5.5pt}
\multicolumn{1}{|c|}{}                          & Average weight                                                                              & \multicolumn{1}{c|}{22.05}     & \multicolumn{1}{c|}{35.86}  & \multicolumn{1}{c|}{25.42}  &                          & \multicolumn{1}{c|}{17.72}     & \multicolumn{1}{c|}{27.17}  & \multicolumn{1}{c|}{20.72}  &                             \\ \hline
\multicolumn{1}{|c|}{\multirow{6}{*}{\rotatebox[origin=c]{90}{Prompt 2}}} & DATETIME  & \multicolumn{1}{c|}{29.62}     & \multicolumn{1}{c|}{56.29}  & \multicolumn{1}{c|}{38.51}  & \multirow{6}{*}{\includegraphics[width=0.20\textwidth]{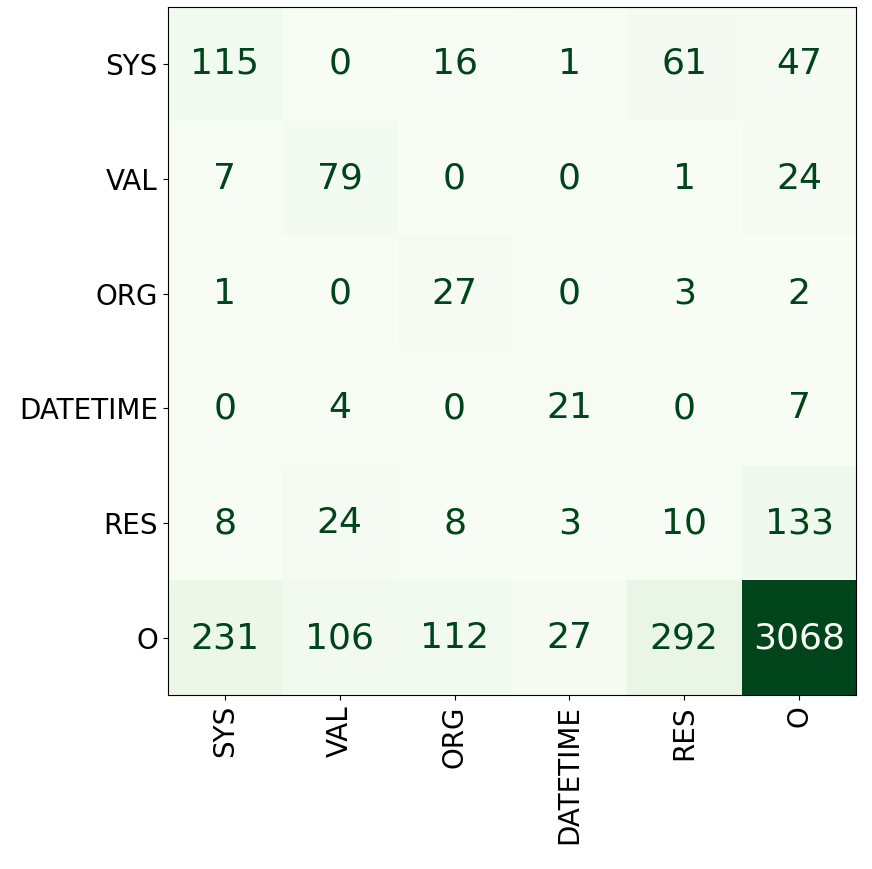}}  & \multicolumn{1}{c|}{48.01}     & \multicolumn{1}{c|}{73.26}  & \multicolumn{1}{c|}{57.298} & \multirow{6}{*}{\includegraphics[width=0.20\textwidth]{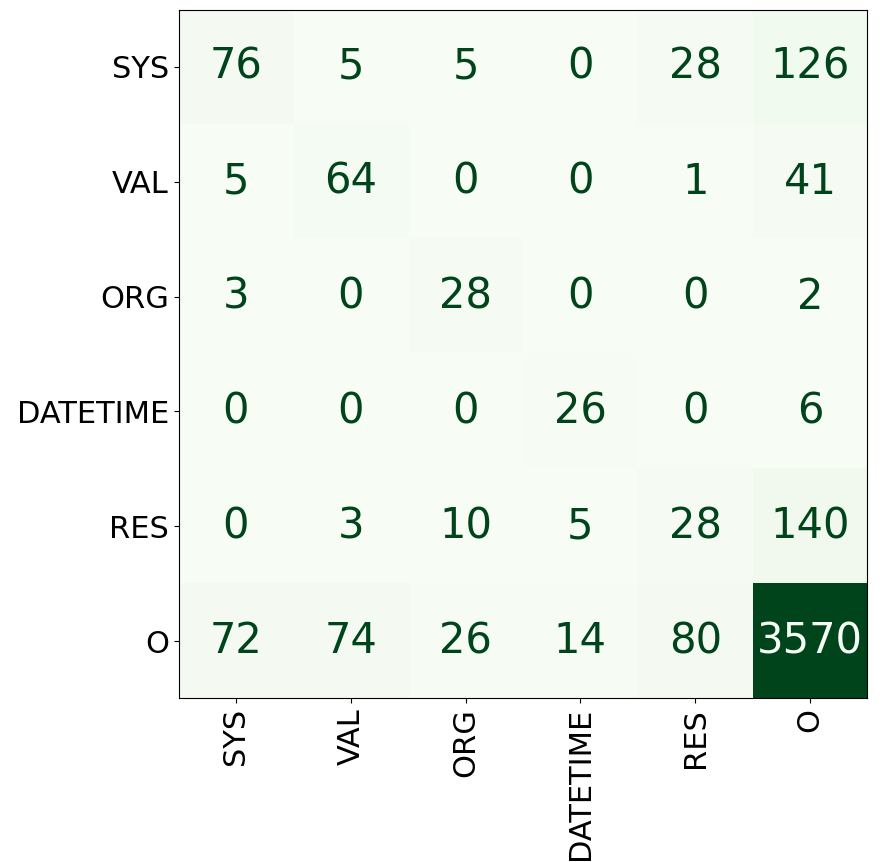}}    \\ \XGap{-2.5pt} \cline{2-5} \cline{7-9} \XGap{-5.5pt}
\multicolumn{1}{|c|}{}                          & ORG       & \multicolumn{1}{c|}{13.21}     & \multicolumn{1}{c|}{61.48}  & \multicolumn{1}{c|}{21.656} &                          & \multicolumn{1}{c|}{44.7}      & \multicolumn{1}{c|}{72.61}  & \multicolumn{1}{c|}{55.188} &                             \\ \XGap{-2.5pt} \cline{2-5} \cline{7-9} \XGap{-5.5pt}
\multicolumn{1}{|c|}{}                          & RES       & \multicolumn{1}{c|}{0.6}       & \multicolumn{1}{c|}{3.28}   & \multicolumn{1}{c|}{1.008}  &                          & \multicolumn{1}{c|}{0.94}      & \multicolumn{1}{c|}{1.78}   & \multicolumn{1}{c|}{1.234}  &                             \\ \XGap{-2.5pt} \cline{2-5} \cline{7-9} \XGap{-5.5pt}
\multicolumn{1}{|c|}{}                          & SYS       & \multicolumn{1}{c|}{0.27}      & \multicolumn{1}{c|}{42.08}  & \multicolumn{1}{c|}{33.068} &                          & \multicolumn{1}{c|}{0.4652}    & \multicolumn{1}{c|}{26.68}  & \multicolumn{1}{c|}{33.989} &                             \\ \XGap{-2.5pt} \cline{2-5} \cline{7-9} \XGap{-5.5pt}
\multicolumn{1}{|c|}{}                          & VAL       & \multicolumn{1}{c|}{17.95}     & \multicolumn{1}{c|}{35.01}  & \multicolumn{1}{c|}{23.574} &                          & \multicolumn{1}{c|}{30.35}     & \multicolumn{1}{c|}{40.5}   & \multicolumn{1}{c|}{34.676} &                             \\ \XGap{-2.5pt} \cline{2-5} \cline{7-9} \XGap{-5.5pt}
\multicolumn{1}{|c|}{}                          & Average weight                                                                              & \multicolumn{1}{c|}{17.75}     & \multicolumn{1}{c|}{39.63}  & \multicolumn{1}{c|}{23.56}  &                          & \multicolumn{1}{c|}{34.1}      & \multicolumn{1}{c|}{42.97}  & \multicolumn{1}{c|}{36.45}  &                             \\ \hline
\multicolumn{1}{|c|}{\multirow{6}{*}{\rotatebox[origin=c]{90}{Prompt 3}}} & DATETIME  & \multicolumn{1}{c|}{39.31}     & \multicolumn{1}{c|}{49.19}  & \multicolumn{1}{c|}{43.302} & \multirow{6}{*}{\includegraphics[width=0.20\textwidth]{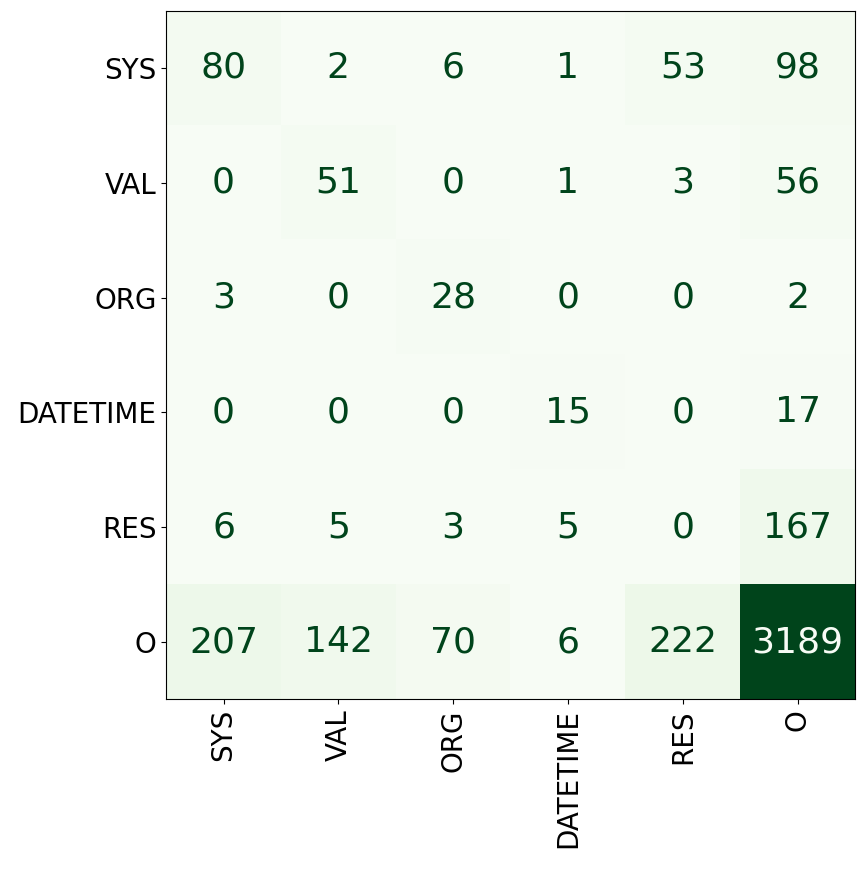}}  & \multicolumn{1}{c|}{50.09}     & \multicolumn{1}{c|}{65.19}  & \multicolumn{1}{c|}{55.82}  & \multirow{6}{*}{{\includegraphics[width=0.20\textwidth]{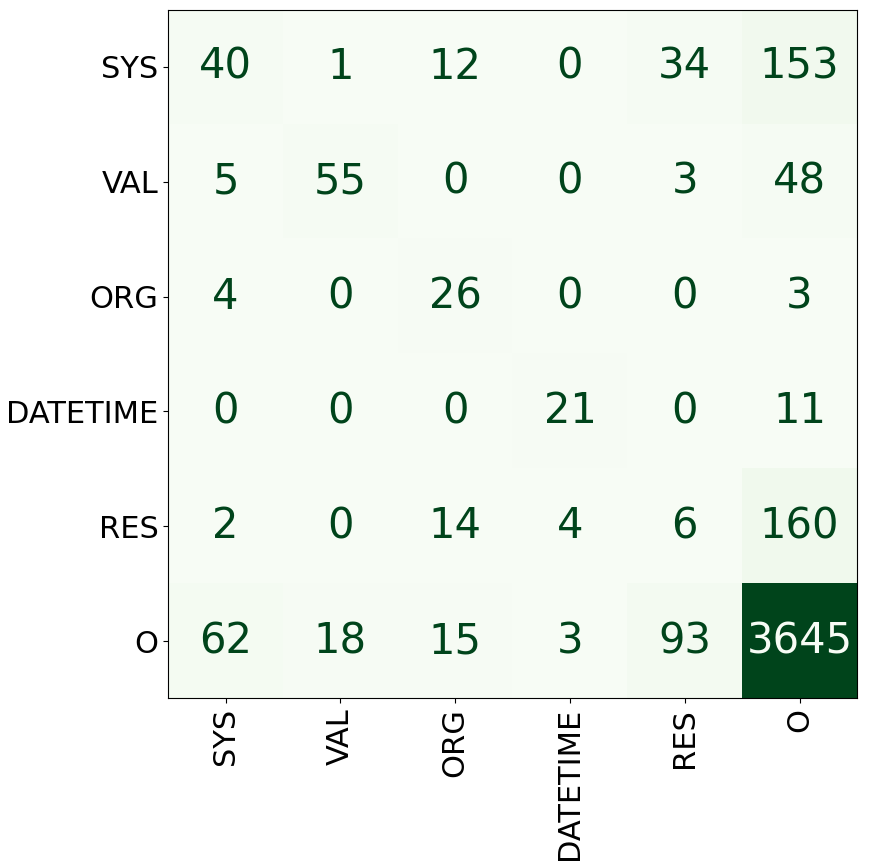}}}    \\ \XGap{-2.5pt} \cline{2-5} \cline{7-9} \XGap{-5.5pt}
\multicolumn{1}{|c|}{}                          & ORG       & \multicolumn{1}{c|}{23.02}     & \multicolumn{1}{c|}{67.98}  & \multicolumn{1}{c|}{34.024} &                          & \multicolumn{1}{c|}{44.84}     & \multicolumn{1}{c|}{75.76}  & \multicolumn{1}{c|}{55.976} &                             \\ \XGap{-2.5pt} \cline{2-5} \cline{7-9} \XGap{-5.5pt}
\multicolumn{1}{|c|}{}                          & RES       & \multicolumn{1}{c|}{0.63}      & \multicolumn{1}{c|}{2.69}   & \multicolumn{1}{c|}{1.018}  &                          & \multicolumn{1}{c|}{1.07}      & \multicolumn{1}{c|}{2.31}   & \multicolumn{1}{c|}{1.45}   &                             \\ \XGap{-2.5pt} \cline{2-5} \cline{7-9} \XGap{-5.5pt}
\multicolumn{1}{|c|}{}                          & SYS       & \multicolumn{1}{c|}{25.22}     & \multicolumn{1}{c|}{32.59}  & \multicolumn{1}{c|}{28.412} &                          & \multicolumn{1}{c|}{30.27}     & \multicolumn{1}{c|}{10.84}  & \multicolumn{1}{c|}{15.852} &                             \\ \XGap{-2.5pt} \cline{2-5} \cline{7-9} \XGap{-5.5pt}
\multicolumn{1}{|c|}{}                          & VAL       & \multicolumn{1}{c|}{13.84}     & \multicolumn{1}{c|}{33.34}  & \multicolumn{1}{c|}{19.546} &                          & \multicolumn{1}{c|}{41.71}     & \multicolumn{1}{c|}{35.08}  & \multicolumn{1}{c|}{37.602} &                             \\ \XGap{-2.5pt} \cline{2-5} \cline{7-9} \XGap{-5.5pt}
\multicolumn{1}{|c|}{}                          & Average weight                                                                              & \multicolumn{1}{c|}{20.4}      & \multicolumn{1}{c|}{37.16}  & \multicolumn{1}{c|}{25.26}  &                          & \multicolumn{1}{c|}{33.59}     & \multicolumn{1}{c|}{37.84}  & \multicolumn{1}{c|}{33.34}  &                             \\ \hline
\end{tabular}
}
\end{table*}

Table \ref{prompts-QA} illustrates performance comparison of GLMs in terms of 9 different evaluation measures across question answering task benchmark dataset. It is evident \deleted{form } \added{from} Table \ref{prompts-QA}, that both models  exhibit varying performance across different prompts. At prompt 1 both GLMS attain low Rouge, BLEU and Meteor score, indicating lack of their capability to capture linguistic features and  significant discrepancies between reference and generated answers. Both GLMS achieve highest Rouge, BLEU and Meteor score at Prompt3 that highlights their ability to capture semantic rich features  and generate highly relevant answers with increased knowledge. A thorough  performance analysis  reveal that ChatGPT surpasses Gemini across all evaluation measures over all 3 prompts .  This indicate that ChatGPT is competent in generating accurate answers by capturing linguistic  features.

\begin{table*}[]
\caption{Generative language models performance comparison with three different prompts across question answering task benchmark datasets}
\label{prompts-QA}
\renewcommand{\arraystretch}{1.5}
\resizebox{1.0\textwidth}{!}{
\begin{tabular}{|l|l|lllllllll|}
\hline
\multirow{2}{*}{\textbf{Models}} & \multirow{2}{*}{\begin{tabular}[c]{@{}l@{}}\textbf{Prompts with Different} \\ \textbf{Expert Knowledge}\end{tabular}} & \multicolumn{9}{c|}{\textbf{Evaluation Measures}}           \\   \XGap{-2.5pt} \cline{3-11} \XGap{-5.5pt} 
&   & \multicolumn{1}{c|}{\textbf{Precision}} & \multicolumn{1}{c|}{\textbf{Recall}} & \multicolumn{1}{c|}{\textbf{F1}}    & \multicolumn{1}{c|}{\textbf{Rouge1}} & \multicolumn{1}{c|}{\textbf{Rouge2}} & \multicolumn{1}{c|}{\textbf{RougeL}} & \multicolumn{1}{c|}{\textbf{RougeLsum}} & \multicolumn{1}{c|}{\textbf{BLEU}}  & \textbf{Meteor} \\ \hline
\multicolumn{1}{|c|}{\multirow{3}{*}{\rotatebox[origin=c]{90}{Gemini}}}  & Prompt 1                  & \multicolumn{1}{c|}{0.822}     & \multicolumn{1}{c|}{0.891}  & \multicolumn{1}{c|}{0.855} & \multicolumn{1}{c|}{0.253}  & \multicolumn{1}{c|}{0.187}  & \multicolumn{1}{c|}{0.236}  & \multicolumn{1}{c|}{0.236}     & \multicolumn{1}{c|}{0.044} & 0.338  \\  \XGap{-2.5pt} \cline{2-11} \XGap{-5.5pt} 
\multicolumn{1}{|c|}{}& Prompt 2                  & \multicolumn{1}{c|}{0.827}     & \multicolumn{1}{c|}{0.898}  & \multicolumn{1}{c|}{0.86}  & \multicolumn{1}{c|}{0.283}  & \multicolumn{1}{c|}{0.213}  & \multicolumn{1}{c|}{0.265}  & \multicolumn{1}{c|}{0.268}     & \multicolumn{1}{c|}{0.056} & 0.377  \\ \XGap{-2.5pt} \cline{2-11} \XGap{-5.5pt}
\multicolumn{1}{|c|}{}& Prompt 3                  & \multicolumn{1}{c|}{0.862}     & \multicolumn{1}{c|}{0.918}  & \multicolumn{1}{c|}{0.888} & \multicolumn{1}{c|}{0.461}  & \multicolumn{1}{c|}{0.41}   & \multicolumn{1}{c|}{0.451}  & \multicolumn{1}{c|}{0.45}      & \multicolumn{1}{c|}{0.187} & 0.567  \\ \hline
\multicolumn{1}{|c|}{\multirow{3}{*}{\rotatebox[origin=c]{90}{ChatGPT}}} & Prompt 1                  & \multicolumn{1}{c|}{0.849}     & \multicolumn{1}{c|}{0.869}  & \multicolumn{1}{c|}{0.858} & \multicolumn{1}{c|}{0.223}  & \multicolumn{1}{c|}{0.096}  & \multicolumn{1}{c|}{0.19}   & \multicolumn{1}{c|}{0.19}      & \multicolumn{1}{c|}{0.055} & 0.232  \\ \XGap{-2.5pt} \cline{2-11} \XGap{-5.5pt} 
\multicolumn{1}{|c|}{}& Prompt 2                  & \multicolumn{1}{c|}{0.888}     & \multicolumn{1}{c|}{0.916}  & \multicolumn{1}{c|}{0.902} & \multicolumn{1}{c|}{0.493}  & \multicolumn{1}{c|}{0.389}  & \multicolumn{1}{c|}{0.464}  & \multicolumn{1}{c|}{0.463}     & \multicolumn{1}{c|}{0.246} & 0.576  \\ \XGap{-2.5pt} \cline{2-11} \XGap{-5.5pt} 
\multicolumn{1}{|c|}{}& Prompt 3                  & \multicolumn{1}{c|}{0.897}     & \multicolumn{1}{c|}{0.929}  & \multicolumn{1}{c|}{0.912} & \multicolumn{1}{c|}{0.566}  & \multicolumn{1}{c|}{0.508}  & \multicolumn{1}{c|}{0.555}  & \multicolumn{1}{c|}{0.556}     & \multicolumn{1}{c|}{0.334} & 0.65   \\ \hline
\end{tabular}}
\end{table*}

\subsubsection{Prompt driven error analysis}
\begin{figure*}[!ht]
  \centering
  \begin{tabular}{cccc}
   & Requirement Extraction &  Requirement Classification & Name Entity Recognition \\
 \rotatebox[origin=c]{90}{ \hspace{1in}Gemini}   & \includegraphics[width=0.30\textwidth]{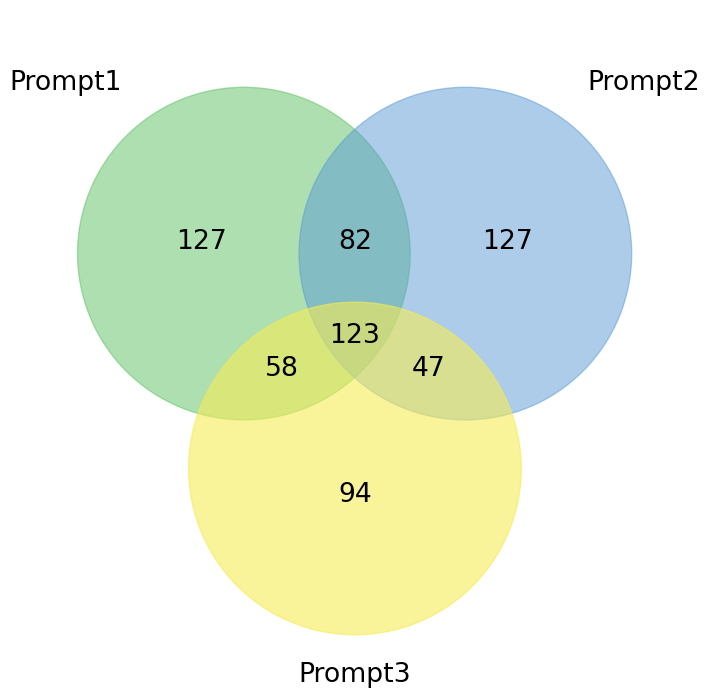}  &  \includegraphics[width=0.30\textwidth]{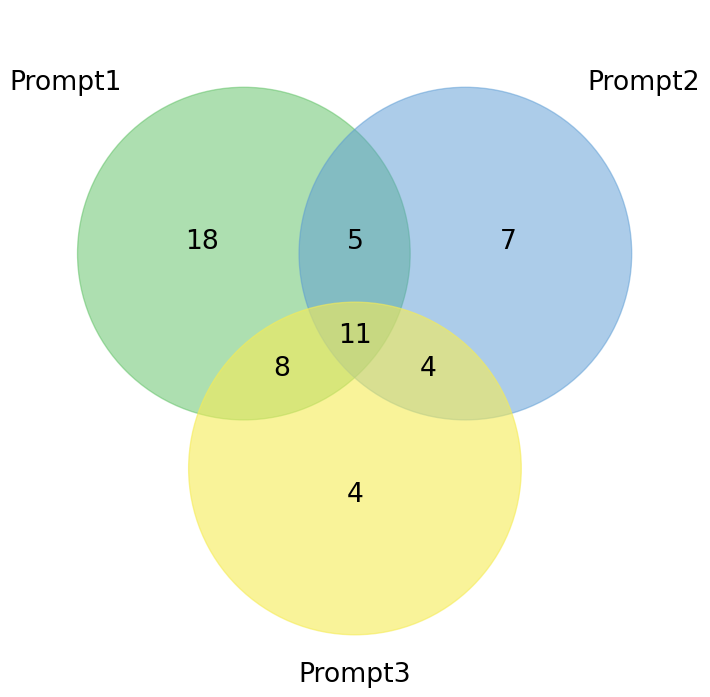} &  \includegraphics[width=0.30\textwidth]{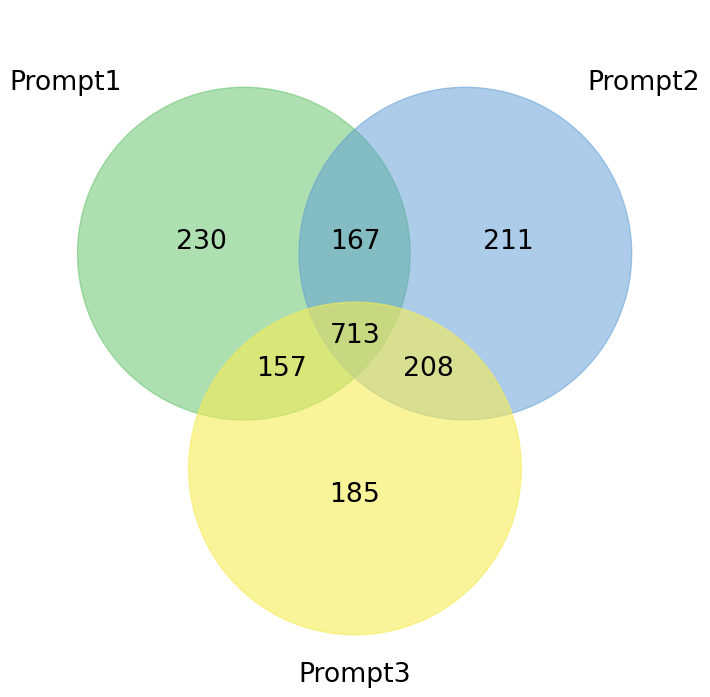}  \\
 & (a) & (b) & (c)
\\[12pt]
  \end{tabular}

\begin{tabular}{cccc}
 \rotatebox[origin=c]{90}{ \hspace{1in}ChatGPT}  & \includegraphics[width=0.30\textwidth]{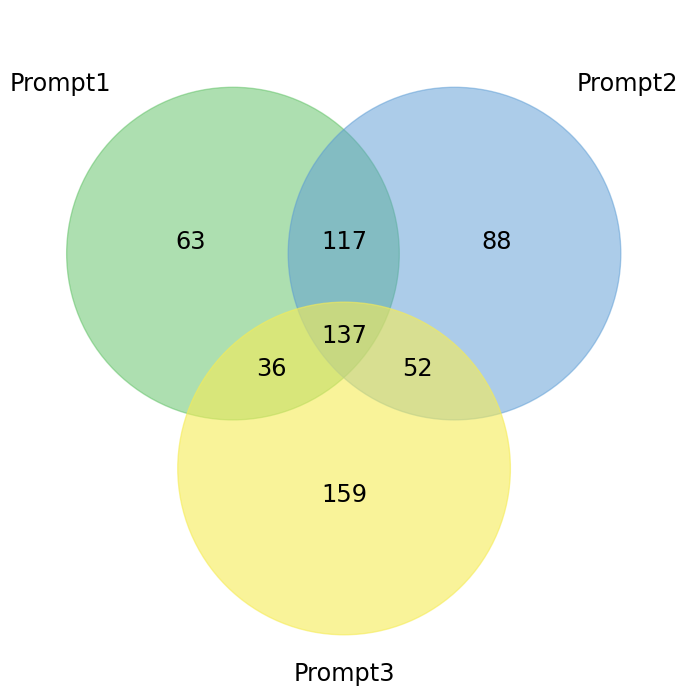}  & \includegraphics[width=0.30\textwidth]{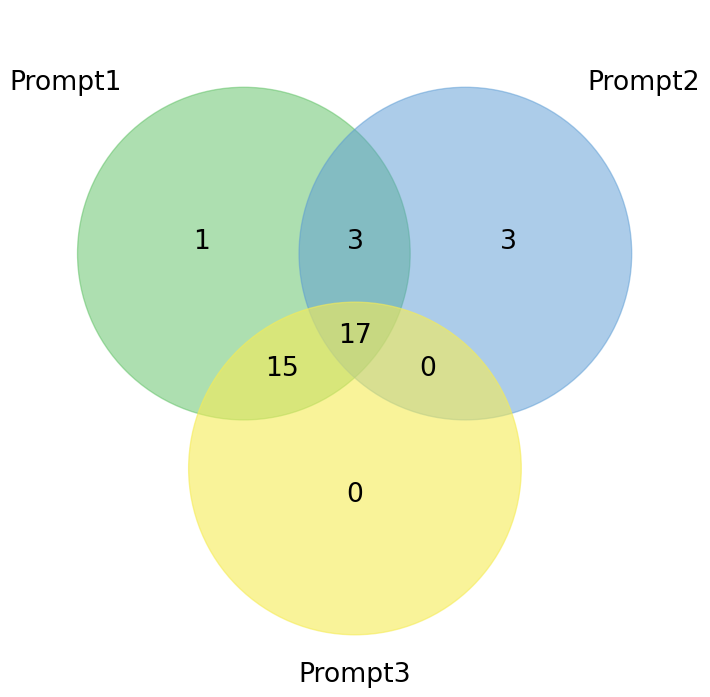} & \includegraphics[width=0.30\textwidth]{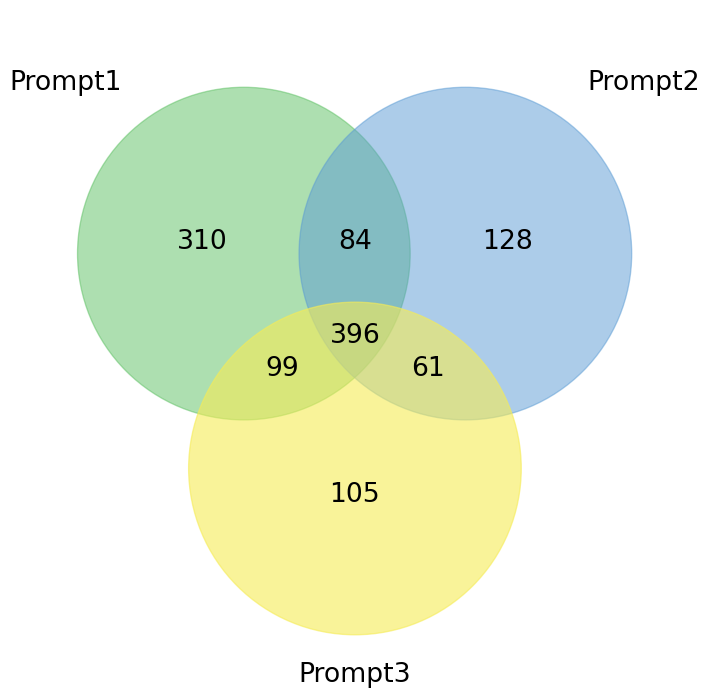} \\
 & (d) & (e) & (f)
  \end{tabular}
  
  \caption{Venn diagrams of generative models across 3 different tasks}
  \label{Ven-diagram}
\end{figure*}

This section briefly illustrates how different prompts impact \added{the} performance of GLMs for different requirement's engineering tasks.  Figure \ref{Ven-diagram} venn diagrams graphically illustrates number of wrong samples predicted by ChatGPT and Gemini using 3 distinct prompts across 3 different tasks (requirement extraction, classification and NER tagging). Specifically, for each language model,  Venn diagram highlights wrong predictions i.e., number of samples wrongly predicted by each prompt and common samples that are wrongly predicted by two and three prompts.  For instance, Figure \ref{Ven-diagram} (a) indicates that  for requirements extraction task, Gemini made 127, 127 and 94 incorrect predictions corresponding to prompts 1, 2, and 3, respectively. The common wrong predictions for prompts 1 and 2 are 82 , for prompts 1 and 3 are 5, for prompts 2 and 3 are 47 and for all three prompts are 123.

A thorough comparative analysis of venn diagrams for both GLMs across requirements extraction and classification tasks reveals that both models exhibit a higher frequency of common errors for prompt pair $p_{1}p_{2}$ and $p_{2}p_{3}$, respectively. This analysis illustrates that  knowledge rich prompts facilitate GLMs requirements extraction task but introduces confusions in requirement classification task for GLMs. Conversely, prompt pairs $p_{2}p_{3}$ and $p_{3}p_{1}$ \deleted{has}\added{have} the highest occurrence of common wrong predictions across NER task for Gemini and ChatGPT, respectively. This analysis highlights that despite having different level of expert knowledge, certain pairs show similar behaviour towards target samples and does not necessarily empower GLMs to discriminate between different classes. Moreover, generally ChatGPT exhibit higher number of common wrong predictions for all 3 prompts as compared to Gemini. which indicates that Gemini's performance is more sensitive to contextual information provided in prompts. 

\subsection{Generative language models performance reproducibility analysis}
To evaluate performance \deleted{reproduce-ability}\added{reproducibility} of GLMs, we selected 30 samples from benchmark dataset of requirements extraction. Specifically,  we obtained five predictions for each sample across all 3 prompts from both GLMs. Across all five prediction iterations, ChatGPT consistently assigned the same class to 66.67\%, 30\%, and 43.34\% samples for prompts 1, 2 and 3, respectively. Contrarily, Gemini consistently predicted the same class for 43.34\%, 53.34\% and 23.34\% samples for prompts 1, 2 and 3, respectively. The lack of consistent predictions across multiple iterations for each prompt highlights a potential limitation in stability of both GLMs. \added{However, since code of these GLMs is not open-source, hence it is challenging to directly addressing the root cause of variability. To enhance reliability, averaging prediction probabilities over multiple iterations can help to stabilize outputs despite the models' closed-source nature.}

\subsection{Generative language models performance comparison with existing predictors}
This section comprehensively illustrates performance comparison of generative language models with existing predictors across four different requirement engineering tasks. It can be summarized from Section \ref{related-work}, that researchers have developed 2 predictors \citep{ivanov2021extracting, sainani2020extracting} for requirements extraction, 20 predictors \citep{toth2019comparative,  haque2019non, baker2019automatic, rahman2019classifying, hey2020norbert, rahimi2020ensemble, dias2020software, tiun2020classification, favero2021bert_se, kici2021bert, rahimi2021one, althanoon2021supporting, kici2021text, ajagbe2022retraining, khayashi2022deep, ivanov2022extracting, kaur2022sabdm, li2022automatic, luo2022prcbert, saleem2023fnreq} for requirements classification, 4 predictors  \citep{ezzini2023ai, abualhaija2022automated, abualhaija2022coreqqa, do2021towards} for QA, and 20 predictors \citep{zhou2018recognizing, pudlitz2019extraction, li2019feature, zhou2020improving, tabassum2020code, lopez2021mining, malik2021named, herwanto2021named, vineetha2022passive, imam2021svm, vineetha2022multinomial, kocerka2022ontology, malik2022identifying, zhou2022named, malik2022software, chow2023analysis, tang2023attensy, malik2023transfer, das2023zero, tikayat2023aerobert} for NER. However, \deleted{among these predictors} most of the\added{se} predictors are evaluated on in-house datasets. This \deleted{paper in hand}\added{study specifically} compares performance of GLMs with \deleted{only those existing predictors whose performance has been reported on the same public benchmark datasets that are used in this study} \added{existing predictors that have been evaluated on the same public benchmark datasets used here. These benchmarks (detailed in Section 3.2) include 13 existing predictors for requirements classification \citep{toth2019comparative,  haque2019non, rahman2019classifying, hey2020norbert,  tiun2020classification,  kici2021bert,  althanoon2021supporting, kici2021text, ajagbe2022retraining, kaur2022sabdm, li2022automatic, luo2022prcbert, saleem2023fnreq}, 1 for requirements extraction \citep{ivanov2021extracting}, 4 for QA \citep{ezzini2023ai, abualhaija2022automated, abualhaija2022coreqqa, do2021towards} and 1 for NER \citep{tikayat2023aerobert}.}\deleted{Across 4 requirement engineering tasks public benchmark datasets (briefly described in section \ref{dataset-section}), \added{the} performance of 13 predictors is reported in the context of classification , 1 predictor for requirements extraction,  4 predictors for QA  and 1 predictor for requirements NER tagging.}

\subsubsection{Generative language models performance comparison with existing requirements extraction predictors}
\begin{figure*}
  \centering
  \includegraphics[width=1.0\textwidth]{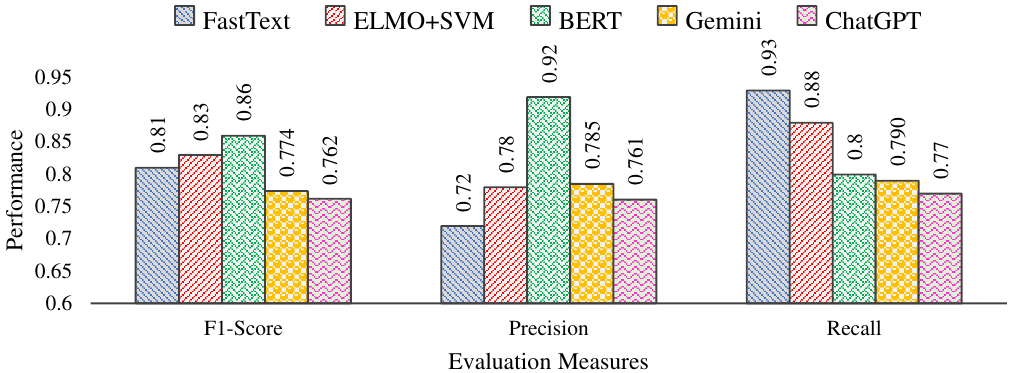} 
  \caption{ Comparative performance analysis of generative language models and existing requirements extraction predictors in terms of three different evaluation metrics}
  \label{req-extraction-SOTA}
\end{figure*}

Figure  \ref{req-extraction-SOTA} illustrates comparative performance analysis of GLMs and 3 existing predictors across  requirements extraction benchmark dataset. Among 3 existing predictors, FastText, ELMo+SVM and BERT managed to produced f1-score of 0.81\%, 0.83\%, and 0.86\%, respectively. FastText predictor remained least performer as it does not utilize any CNN, RNN, or attention mechanism to extract useful features. It  initializes random word embeddings and during training phase it updates embedding weights. Contrarily, ELMo+SVM predictor managed to produced better performance as it leverages the contextual embeddings of ELMo (Embeddings from language models) along with discriminative potential of SVM classifier. Bidirectional contextual embeddings facilitates BERT to capture semantic information; outperforming other two existing predictors. Furthermore, ChatGPT and Gemini produced f1-score of 76\% and 79\%, respectively\added{,} and \deleted{remain} fail to outperform existing predictors,

\subsubsection{Generative language models performance comparison with existing requirements classification predictors}

\begin{figure*}{}
  \centering
  \includegraphics[width=1.0\textwidth]{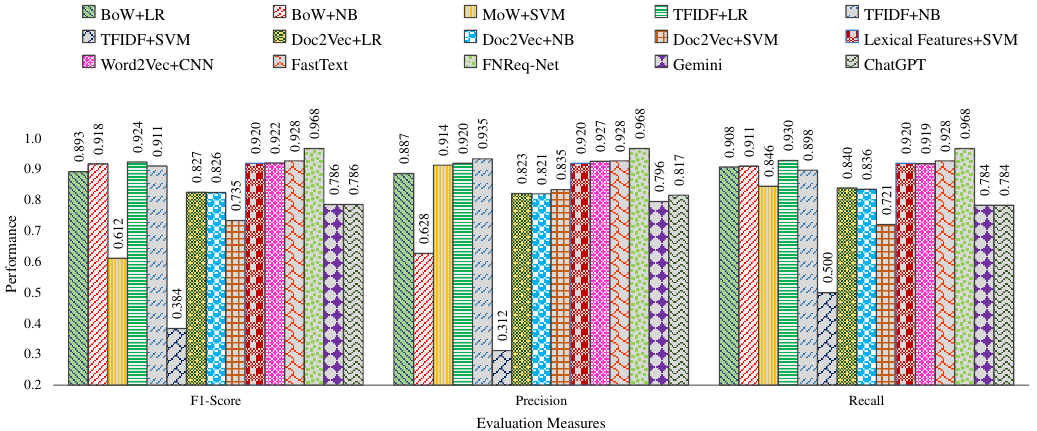} 
  \caption{Comparative performance analysis of generative language models and existing requirements classification predictors in terms of three different evaluation metrics}
  \label{req-classification-SOTA}
\end{figure*}

Figure  \ref{req-classification-SOTA}  graphically represents performance comparison of GLMs and 13 existing predictors \citep{toth2019comparative,  haque2019non, rahman2019classifying, hey2020norbert,  tiun2020classification,  kici2021bert,  althanoon2021supporting, kici2021text, ajagbe2022retraining, kaur2022sabdm, li2022automatic, luo2022prcbert, saleem2023fnreq}, across requirements classification benchmark dataset in terms of three distinct evaluation metrics. Among 13 existing requirements classification predictors, bag of words or TFIDF based representation with SVM classifier produced least performance \citep{dias2020software}. \added{The traditional predictor struggle to capture comprehensive semantic relationships of complex requirements \citep{dias2020software}.} \deleted{Unlike 2 aforementioned predictive pipelines, a slightly better performance is produced by three predictive pipelines that utilized  Doc2Vec feature representation strategy with SVM, NB, or LR classifiers.} \added{In contrast, predictors utilizing Word2Vec and FastText embeddings showed better performance, highlighting that advanced embedding techniques provide a richer representation of requirements. Among all predictors, FnReq has state-of-the-art performance as its predictive pipeline is enriched with feature selection method and attention architecture to facilitate more discriminative features. } The GLMs managed to outperform only three existing predictive pipelines namely; BoW + SVM, TFIDF + SVM and Doc2Vec + SVM. This outcome suggests that while GLMs excel in many NLP tasks, their general training may not fully capture the specialized nature of requirements text. 
\deleted{Moreover, as compared to previously mentioned predictive pipelines, a significant performance boost is achieved by 3 different predictive pipelines namely: BoW and  TFIDF based representation technique  with NB, BoW +  LR and TFIDF + LR. A slight better performance is produced by four different predictive pipelines including: TFIDF + LR, Word2Vec + CNN, lexical features + SVM classifier and FastText. FnReq managed to produce \added{the} highest performance.}

\subsubsection{Generative language models performance comparison with existing requirements NER tagger}

\begin{table*}[]
\caption{Comparative performance analysis of generative language models and existing requirements based NER predictor across five distinct NER tags}
\label{NER-SOTA}
\renewcommand{\arraystretch}{1.5}
\resizebox{1.0\textwidth}{!}{
\begin{tabular}{|c|ccc|ccc|ccc|}
\hline
\multicolumn{1}{|l|}{\multirow{2}{*}{NER Tag}} & \multicolumn{3}{c|}{Aero-BERT} & \multicolumn{3}{c|}{ChatGPT}                                          & \multicolumn{3}{c|}{Gemini}                                          \\  \XGap{-2.5pt} \cline{2-10} \XGap{-5.5pt}
\multicolumn{1}{|l|}{}                            & Precision   & Recall   & F1    & \multicolumn{1}{c|}{Precision} & \multicolumn{1}{c|}{Recall} & F1     & \multicolumn{1}{c|}{Precision} & \multicolumn{1}{c|}{Recall} & F1    \\
DATETIME                                          & 0.88        & 0.88     & 0.88  & \multicolumn{1}{c|}{0.48}     & \multicolumn{1}{c|}{0.73}  & 0.57 & \multicolumn{1}{c|}{0.43}     & \multicolumn{1}{c|}{0.46}  & 0.43 \\
ORG                                               & 0.98        & 0.92     & 0.95  & \multicolumn{1}{c|}{0.44}     & \multicolumn{1}{c|}{0.72}  & 0.55 & \multicolumn{1}{c|}{0.27}      & \multicolumn{1}{c|}{0.74}  & 0.40 \\
RES                                               & 0.98        & 0.98     & 0.98  & \multicolumn{1}{c|}{0.94}      & \multicolumn{1}{c|}{0.01}   & 0.012  & \multicolumn{1}{c|}{0.019}      & \multicolumn{1}{c|}{0.03}   & 0.018  \\
SYS                                               & 0.93        & 0.91     & 0.92  & \multicolumn{1}{c|}{0.46}    & \multicolumn{1}{c|}{0.26}  & 0.33 & \multicolumn{1}{c|}{0.29}     & \multicolumn{1}{c|}{0.28}  & 0.28 \\
VAL                                               & 0.89        & 0.91     & 0.9   & \multicolumn{1}{c|}{0.30}     & \multicolumn{1}{c|}{0.40}   & 0.34 & \multicolumn{1}{c|}{0.08}      & \multicolumn{1}{c|}{0.26}  & 0.12 \\
Weighted average                                  & 0.93        & 0.92     & 0.92  & \multicolumn{1}{c|}{0.34}      & \multicolumn{1}{c|}{0.42}  & 0.36  & \multicolumn{1}{c|}{0.22}     & \multicolumn{1}{c|}{0.35}  & 0.25 \\ \hline
\end{tabular}}
\end{table*}
Table \ref{NER-SOTA} presents performance comparison of existing predictor and GLMs across 3 different evaluation measures. Existing predictor Aero-BERT consistently excels across all measures to recognize all five entity types i.e. DATETIME, ORG, RES, SYS and VAL. Furthermore, it also achieves a great balance between precision and recall across all NER types. In contrast, GLMs lack to maintain this balance and struggles with imbalanced trade-off between precision and recall. Aero-BERT consistently outperforms both GLMs in terms of weighted average across across all entity types.

\subsubsection{Generative language models performance comparison with existing requirements specific question answering systems}
\begin{figure*}
  \centering
  \includegraphics[width=1.0\textwidth]{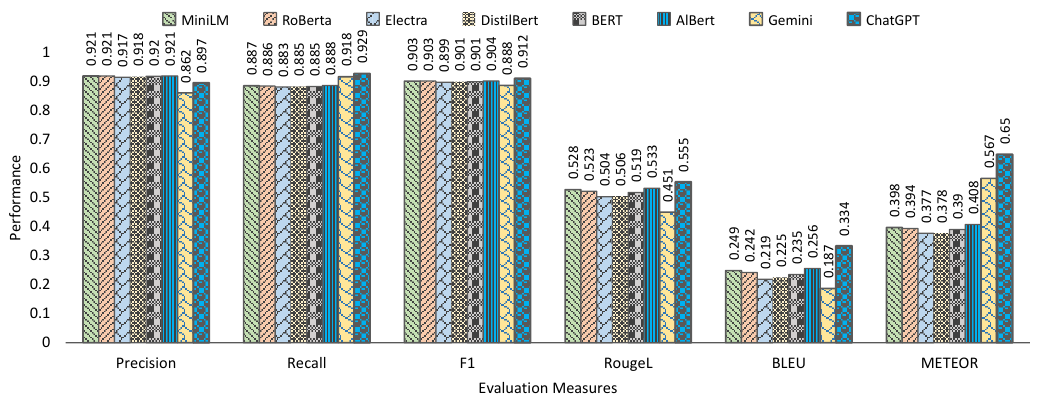} 
  \caption{Comparative performance analysis of generative language models and existing requirements
based question answering systems}
\label{QA-existing-results}
\end{figure*}

Figure \ref{QA-existing-results} provides performance comparison of existing predictors and GLMs across 6 different evaluation measures for benchmark dataset of requirements-based question-answering. A thorough \deleted{comparative} analysis of Figure \ref{QA-existing-results} reveals that among 6 existing predictors \citep{ezzini2023ai}, Electra , DistilBERT, and BERT achieve slightly lower results and secures \added{rank at} $6^{th}$, $5^{th}$  and $4^{th}$ position respectively. RoBERTa  secures $3^{rd}$ position and MiniLM follows closely at $2^{nd}$ rank.  Albert beats all other  existing predictors which indicates its superior ability to capture semantic information. Both GLMs outshine top-performing AlBert model across various metrics which demonstrate their effectiveness in comparison to other generic language models. Furthermore, comparative performance analysis of GLMs reveal that ChatGPT surpasses Gemini in terms of recall, BLEU, METEOR, ROUGE score indicates ChatGPT's superior ability to capture relevant information.

\section{Practical implication}
The recent surge in the development of LLMs has empowered  software development life cycle by replacing manual works with computer aided applications \citep{hey2020norbert, luo2022prcbert, ivanov2021extracting}. Potential of LLMs has been explored in diverse types of scientific studies across distinct types of requirement engineering tasks  \citep{alturayeif2023automated, luo2022prcbert, ivanov2021extracting}. Specifically, potential of BERT language model is harnessed to develop requirements extraction and classification applications. Similarly, BERT \citep{devlin2018bert} and T5 \citep{lin2019commongen} language models potential is explored for NER tagging of requirements. For requirements question answering task, researchers have explored the potential of 6 different language models including; RoBerta \citep{liu2019roberta}, Electra \citep{clark2020electra}, DistilBERT \citep{sanh2019distilbert}, MiniLM \citep{wang2020minilm}, Albert \citep{lan2019albert} and BERT \citep{devlin2018bert}. On the other hand, the most recent GLMs including; ChatGPT \citep{radford2019language} and Gemini \citep{pal2024gemini} are considered more useful tools for development of NLP applications. However, there is lack of scientific studies \citep{yeow2024automated, wu2024chatgpt, hamdi2023prompt, alter2024could} having focus on exploration of GLMs potential in software systems field. Primary objective of this study is to analyze whether GLMs are capable of developing useful requirement engineering applications. Moreover, another objective is to explore the most effective methods for enhancing prompt-based predictions. A thorough experimentation reveals that prompts with more domain specific keywords tend to introduce biassness towards a particular class for classification tasks. Similarly, for NER task  both GLMs  showed best performance with  prompts having less domain specific keywords. This analysis concludes that for both classification and NER tagging tasks, basic prompts outperform knowledge rich prompts. In other words, using prompts that are more broadly applicable leads to better results.

In question-answering task, \added{the} performance of both GLMs  improved significantly when they are fed with domain-specific knowledge alongside prompts containing questions. In contrast, they produce less performance when they are fed with only prompts and are supposed to provide answer based on their background knowledge. Since, these models are trained on large corpora and encompass\deleted{es} only high level insights about different fields. That is why \deleted{they remain}\added{still} fail to acquire domain specific answers of questions from their background knowledge. However, their background knowledge is sufficient for extracting desired answers from a large textual files. Overall, we believe these models are useful for question answering task and may be used for classification tasks with more optimal prompts. The optimal prompts can be designed by employing prompts optimization techniques.

Primarily, prompts are categorized into two main classes namely; hard prompts \citep{zhang2024zero} and soft prompts \citep{wu2022adversarial}. Former encompass discrete human-readable words or instructions while later refers to continuous statistical vectors that are not directly understandable by human. In terms of optimization methods both prompts differ in several aspect. Hard prompts are typically optimized using reinforcement learning, evolutionary algorithms, and gradient-free techniques, including: Auto-CoT \citep{zhang2022automatic},  EPR \citep{rubin2021learning}, LMaaS \citep{sun2022black}, PROPANE \citep{melamed2023propane}, CLAPS, FEDBPT, GRIPS, etc. In contrast, soft prompts rely on standard continuous optimization methods like gradient descent, such as Ptuning \citep{liu2023gpt}, prefix tuning \citep{li2021prefix}, prompt tuning \citep{lester2021power}, P Tuning Version 2 \citep{liu2021p}, etc.

Moreover, hard prompts remain interpretable throughout the optimization process and can be manually edited and refined by humans after optimization. On the other hand, soft prompts are less interpretable. Hard prompts  can be easily transferred between different models and used in text-based interfaces. In contrast, soft prompts tend to be model-specific and cannot be directly applied in text interfaces. Generally soft prompts have been associated with better performance due to their continuous nature. However, recent advancements in hard prompt optimization techniques are narrowing this performance gap. Ultimately, hard prompts are more suitable for applications that require human interpretation or interaction, while soft prompts are often preferred in scenarios where maximizing model performance is the primary objective, regardless of human interpretability. A \deleted{compiling}\added{compelling} future direction of this work is to utilize distinct hard and soft prompt optimization techniques for distinct requirement engineering tasks including requirements extraction, classification, NER tagging and question answering system.

\section{Conclusion}
This research investigates the potential of generative language models across four distinct requirement engineering tasks by employing prompts with varying levels of expert knowledge at three different stages. A comprehensive performance analysis of generative language models across four distinct requirement engineering tasks public benchmark datasets namely, requirement extraction, classification, named entity recognition, and question answering reveals a common trend where the models demonstrate improved performance with an escalation in the level of expert knowledge information provided in prompts. However, this improvement is accompanied by an increased bias toward type 1 or type 2 errors. This analysis illustrates that the performance of models can be directed towards a specific class by crafting prompts with domain-specific keywords. Furthermore, among both models Gemini requires domain specific knowledge enriched prompts to achieve its full potential, while ChatGPT shows greater robustness with less domain specific knowledge based  prompts. In the realm of requirement engineering four tasks, both models exhibit comparable performance in the requirements classification task. \deleted{However, in the}\added{In the} remaining three tasks, ChatGPT consistently outperforms Gemini. \deleted{Moreover, when compared}\added{In comparison} to existing task-specific machine learning models and traditional language model-based predictors, ChatGPT achieves state-of-the-art performance in the question answering task. \deleted{However, for the} \added{For} other three tasks, the performance of existing predictors is significantly better than ChatGPT.
In summary, the findings of this case study suggest that generative language models prove beneficial for question answering tasks. However, there remains a need to harness the potential of traditional machine/deep learning models or foundational language models for applications in requirement engineering.

\section*{Compliance with ethical standards}
\subsection*{Funding} Not applicable.

\subsection*{Conflict of Interest} Corresponding author on the behalf of all authors
declares that no conflict of interest is present.

\bibliographystyle{sn-mathphys} 
\bibliography{sn-article}
\end{document}


\section*{Supplementary Tables}

\begin{table}[h!]
\caption{3 Different Expert level knowledge Prompts for Requirement Extraction}
\label{RE-prompts-table}
\renewcommand{\arraystretch}{0.89}
\resizebox{1.0\textwidth}{!}{
\begin{tabular}{|l|l|}
\hline
\begin{tabular}[c]{@{}l@{}}Expert\\ level 1 \\ knowledge\\ Prompt\end{tabular} & \begin{tabular}[c]{@{}l@{}}In Software Requirement Specification (SRS) of softwares, content belongs\\ to two different classes  namely requirements and non-requirements. \\ Requirements related content contain information like specific functionalities,\\ qualities and characteristics of a software such as features, performance, \\ security, reliability, usability, recoverability, privacy etc. Non-requirements \\ class related content comprises any generic information about software that \\ are not part of core functionalities of software. Based on your background \\ knowledge and aforementioned definitions, you are supposed to categorize\\ content into requirements or non requirements classes: The content you are\\ supposed to classify is: \textless{}text\textgreater{}.\end{tabular}                                                                                                                                                                                                                                                                                                                                                                                                                                                                                                                                                                                                                                                                                                                                                                                                                                    \\ \hline
\begin{tabular}[c]{@{}l@{}}Expert\\ level 2\\ knowledge\\ Prompt\end{tabular}  & \begin{tabular}[c]{@{}l@{}}In Software Requirement Specification (SRS) document, content belongs\\ to two different classes namely requirements and non-requirements. \\ Requirements related content contain information like specific functionalities,\\ qualities and characteristics of a software such as features, performance,\\ security, reliability, usability, recoverability, privacy etc. For instance, it can\\ specify that a word processing software must include features like text \\ formatting, spell checking, and printing capabilities. Non-requirements class\\ related content comprises any generic information about software that are not\\ part of core functionalities of software. Based on your background knowledge\\ and aforementioned definitions, you are suppose to categorize content into \\ requirements or  non requirements classes: The content you are supposed to \\ classify is: \textless{}text\textgreater{}.\end{tabular}                                                                                                                                                                                                                                                                                                                                                                                                                                                                                                                                                                                                                                                                                \\ \hline
\begin{tabular}[c]{@{}l@{}}Expert\\ level 3\\ knowledge\\ Prompt\end{tabular}  & \begin{tabular}[c]{@{}l@{}}In the field Software Engineering, a Software Requirement Specification\\ (SRS) document serves as blue print to understand  software behavior and\\ user  expectations. However, content of SRS can be  categorized into two\\ distinct classes: requirements and non-requirements.  Requirements class\\ comprises of content that  provides explicit details about what the software\\ is expected to accomplish. This encompasses a variety of elements  \\ including specific features,performance metrics, security standards, reliability\\ expectations, usability guidelines, recoverability processes,privacy etc. For\\ instance, a SRS for a word processing software might state that the software\\ should offer functionalities like text formatting, spell checking, and printing.\\ It may further specify performance expectations such as loading a  document\\ under two seconds. On the other hand, non-requirements class contains \\ content that,  doesn't directly demonstrate functional behavior of software.\\ It might include market tools analysis, potential usage,  generic  information\\ and any other miscellaneous details that don't define the core  functionality\\ of the software. For instance: A section in the SRS describing the features\\ of  existing word processing tools or  would be considered non-requirements. \\ Based on your background knowledge and aforementioned definitions, you \\ are suppose to categorize content into requirements or non requirements \\ classes: The content you are supposed to classify is: \textless{}text\textgreater{}.\end{tabular} \\ \hline
\end{tabular}}
\end{table}

\begin{table}[h!]
\caption{3 Different Expert level knowledge Prompts for Requirement Classification}
\label{RC-prompts-table}
\renewcommand{\arraystretch}{1.3}
\resizebox{1.0\textwidth}{!}{
\begin{tabular}{|l|l|}
\hline
\begin{tabular}[c]{@{}l@{}}Expert\\ level 1 \\ knowledge\\ Prompt\end{tabular} & \begin{tabular}[c]{@{}l@{}}Software requirements can be categorized into functional and non-functional\\ classes. Functional class related requirements define specific features and\\ functionalities of software. Non-functional class represents requirements\\ that define qualities and characteristics of a software. Based on your\\ background knowledge and aforementioned definitions, you are supposed\\ to categorize requirement into functional or non-functional class: The\\ requirement you are supposed to classify is: \textless{}text\textgreater{}\end{tabular}                                                                                                                                                                                                                                                                                                                                                                                                                                                                                                                                                                                                                                                                                                                                                                                                                                                     \\ \hline
\begin{tabular}[c]{@{}l@{}}Expert\\ level 2\\ knowledge\\ Prompt\end{tabular}  & \begin{tabular}[c]{@{}l@{}}Software requirements can be categorized into functional and non-functional \\ classes. Functional class related requirements define specific features and\\  functionalities of software. Non-functional class represents requirements\\  that define qualities and characteristics of a software such as database,\\  security, look and feel, legal, privacy, usability, scalibility, availability,\\  recoverability, reliability. Based on your background knowledge and\\  aforementioned definitions, you are suppose to categorize requirement\\  into functional and non-functional classes: The requirement you are \\ supposed to classify is: \textless{}text\textgreater{}\end{tabular}                                                                                                                                                                                                                                                                                                                                                                                                                                                                                                                                                                                                                                                                                                             \\ \hline
\begin{tabular}[c]{@{}l@{}}Expert\\ level 3\\ knowledge\\ Prompt\end{tabular}  & \begin{tabular}[c]{@{}l@{}}Software requirements play a crucial role in determining the overall design\\  and functionality of a system. Broadly, these requirements can be divided\\  into two main categories: functional and non-functional. Functional\\  Requirements are directly related to the specific functionalities and\\  features of the software. They outline what a system is supposed to do.\\  For example, the ability of a software to generate reports, perform\\  calculations, or display specific information can be classified under\\  functional requirements. Non-functional requirements define the\\  qualities and characteristics a software should possess. Some of the\\  key non-functional requirements include: database requirements\\  (type and structure of database used), security protocols, aesthetic\\  aspects (such as interface), legal and regulatory compliance, privacy\\  measures, usability aspects, scalability (how well the software can handle\\  growth), availability (reliability of software), recoverability (how the\\  software can recover from crashes), reliability (dependability of software).\\  Based on your background knowledge and aforementioned definitions,\\  you are supposed to categorize requirements into functional and \\ non-functional classes: The requirements you are supposed to classify\\  is:  \textless{}text\textgreater{}\end{tabular} \\ \hline
\end{tabular}}
\end{table}

\begin{table}[h!]
\centering
\caption{Different expert level knowledge prompt for requirements NER tagging}
\label{NER-prompts-table}
\renewcommand{\arraystretch}{0.99}
\resizebox{1\textwidth}{!}{
\begin{tabular}{|l|l|}
\hline
\rotatebox[origin=c]{90}{\begin{tabular}[c]{@{}c@{}}Expert level 1 \\ Knowledge Prompt\end{tabular}} &  \begin{tabular}[c]{@{}l@{}}While developing Artificial Intelligence based software engineering applications, Named Entity Recognition (NER)\\ is an important task for extraction of useful  information.  Prime  objective of NER task is to annotate requirements \\content  with predefined set of tags where  each tag represents a unique type of information. Based on your \\background knowledge and  aforementioned information, you are supposed to annotate requirements words with \\predefined name entity tags namely O (Other), B-SYS (Beginning of a System), I-SYS (Inside a System), B-VAL \\(Beginning of a Value),  I-VAL (Inside a Value),  B-DATETIME (Beginning of a Date-Time), I-DATETIME\\ (Inside a Date-Time),   B-ORG (Beginning of an Organization), I-ORG (Inside an Organization),  B-RES\\ (Beginning of  a Resource), and I-RES (Inside a Resource). The content  you are supposed to tag is: \textless{}text\textgreater{}. \\ Please specify your answer in BIO tagging  scheme.  \end{tabular}                                                                                                                                                                                                                                                                                                                                                                                                                                                                                                                                                                                                                                                                                                                                                                                                                                                                                                                                                                                                                                                                                                                                                                                                                                                                                                                                                                                                                                                                                                                                                                                                                                                                                                                                                                                                                                                                                                                                                                                                                                                                                                                                                                                                                                                                                                                                                                                                                                                                                                            \\ \hline
\rotatebox[origin=c]{90}{Expert level 2 knowledge Prompt}  & \begin{tabular}[c]{@{}l@{}}While developing Artificial Intelligence based software engineering applications, Named Entity Recognition\\ (NER) is an important task for extraction of useful  information.  Prime  objective of NER task is to annotate \\requirements content  with predefined set  of tags including O, B-SYS, I-SYS, B-VAL, I-VAL,  B-DATETIME, \\I-DATETIME, B-ORG, I-ORG, B-RES, and I-RES.  Each tag  represents a unique type of information. To \\ understand which tag represents  which type of information, here we provide definitions of tags: 1) Beginning \\ of a System (B-SYS): B-SYS defines starting word of a system or  platform name.  2) Inside a System (I-SYS): \\ I-SYS represents a system or platform name middle words. 3) Beginning of a Value (B-VAL): B-VAL defines\\ values  or quantity of software or hardware components. 4) Inside a Value (I-VAL): I-VAL indicates words\\ about description of values. 5) Beginning of a Date-Time  (B-DATETIME): B-DATETIME defines the start  \\of a date or time.6) Inside a Date-Time (I-DATETIME): I-DATETIME refers to words within a date-time \\ description. 7) Beginning of an Organization (B-ORG): B-ORG defines starting word of an organization name.\\  8) Inside an Organization (I-ORG): I-ORG  represents an organization name middle words. 9) Beginning of \\  a Resource (B-RES): B-RES defines starting words about description of hardware components. 10) Inside a\\ Resource  (I-RES): I-RES represents middle words  about description of hardware components. 10) Other (O): \\O refers to words that does not fall under the paradigm of aforementioned 10 different categories.  Based on \\your  background knowledge and aforementioned tags definitions, you are supposed to annotate requirements \\ words with predefined name entity  tags namely O, B-SYS, I-SYS, B-VAL, I-VAL, B-DATETIME, \\I-DATETIME,  B-ORG, I-ORG, B-RES, and I-RES. The content you are supposed to tag is:\\ \textless{}text\textgreater{}.\end{tabular}                                                                                                                                                                                                                                                                                                                                                                                                                                                                                                                                                                                                                                                                                                                                                                                                                                                                                                                                                                                                                                                                                                                                                                                                                                                                                                                                                                                                                                                                                                               \\ \hline
\rotatebox[origin=c]{90}{Expert level 3 knowledge Prompt} & \begin{tabular}[c]{@{}l@{}}While developing Artificial Intelligence based software engineering applications,  Named Entity Recognition\\ (NER) is an important task for extraction of useful  information.  Prime objective of NER task is to annotate \\requirements content  with predefined set of tags including O, B-SYS, I-SYS, B-VAL, I-VAL, B-DATETIME,\\ I-DATETIME, B-ORG, I-ORG, B-RES, and I-RES. Each tag  represents a unique type of information. To \\understand which tag represents which type of information, here we define some sample requirements.\\ R1: The application should be compatible with the Android operating system.\\ R2: The release date is set for January 15, 2024. \\ R3: The system requires a minimum of 8GB RAM. \\ R4: Google Inc. is the primary stakeholder for this project. \\ R5: The database should be stored on an SSD drive. \\ Definitions of tags along with examples from above sample requirements are\\ defined below: \\ 1) Beginning of a System (B-SYS): B-SYS defines starting word of a system  or platform name. For instance, \\in sample requirement R1, the word "Android" would be tagged as "B-SYS". 2)  Inside a System (I-SYS):\\  I-SYS represents a system or platform name middle words. For instance, in sample requirement R1,"operating" \\would be tagged as "I-SYS".3) Beginning of a Value (B-VAL):  B-VAL defines values or quantity of software or\\ hardware components. For  instance, in sample requirement R3, "8GB" would be tagged as "B-VAL". 4) \\  Inside a Value (I-VAL): I-VAL indicates words about description of values. For  instance, if sample\\ requirement R3 is modified to state"8 gigabytes of RAM",  the word "gigabytes" would be tagged as "I-VAL". \\ 5) Beginning of a Date-Time (B-DATETIME): B-DATETIME defines the start of a date or time. For instance,\\ in sample requirement R2, "January 15" would be tagged as "B-DATETIME". 6) Inside a Date-Time \\(I-DATETIME): I-DATETIME refers to words within a date-time description. For instance, in sample \\requirement R2, "2024" would  be tagged as "I-DATETIME". 7)Beginning of an Organization (B-ORG):\\  B-ORG defines starting word of an organization name. For instance, in sample requirement R4, "Google" \\would be tagged as "B-ORG". 8) Inside an  Organization (I-ORG): I-ORG represents an organization\\ name middle words. For instance, in sample requirement R4, "Inc."  would be tagged as "I-ORG". 9) \\Beginning of a Resource (B-RES): B-RES defines starting words about description of hardware components.\\ For instance, in sample requirement R5, "SSD" would be tagged as "B-RES". 10)  Inside a Resource (I-RES):\\ I-RES represents middle words about description of hardware components. For instance, in sample requirement\\ R5 is modified to provide information about specific type or model of SSD mentioned in R5, the descriptor\\ might be tagged as "I-RES". 11) Other (O): O refers to words that does not fall under the paradigm\\ of aforementioned 10 different categories. For instance, in sample requirement R2, words like "is", "set", and\\ ``for" would be tagged as "O".Based on your  background knowledge and aforementioned tags definitions, you \\are supposed to annotate requirements words with predefined name entity tags namely O, B-SYS, I-SYS,\\ B-VAL, I-VAL, B-DATETIME, I-DATETIME, B-ORG, I-ORG, B-RES, and I-RES. The content you are \\supposed to tag is:\\  \textless{}text\textgreater{}.\end{tabular} \\ \hline
\end{tabular}}
\end{table}

\begin{table}[h]
\caption{Different expert level knowledge prompt for requirements question answering system}
\label{QA-prompts-table}
\renewcommand{\arraystretch}{0.9}
\resizebox{1\textwidth}{!}{
\begin{tabular}{|l|l|}
\hline
\begin{tabular}[c]{@{}l@{}}Expert\\ level 1 \\ knowledge\\ Prompt\end{tabular} & \begin{tabular}[c]{@{}l@{}}Prompt 1 =
``Provide the answer of the given question in \textless{}text\textgreater based on given \\ context."\\ text = Prompt 1 + context{[}i{]} + ``\textless{}" + question{[}i{]} + ``\textgreater{}"\end{tabular}                                                                                                                                        \\ \hline
\begin{tabular}[c]{@{}l@{}}Expert\\ level 2\\ knowledge\\ Prompt\end{tabular}  & \begin{tabular}[c]{@{}l@{}}Prompt 2= ``Provide the answer of the given question in \textless{}text\textgreater{}."\\ text = Prompt 2 + ``\textless{}" + question{[}i{]} + ``\textgreater : " + ``{[}" + context{[}i{]} + ``{]}"\end{tabular}                                                                                                                                              \\ \hline
\begin{tabular}[c]{@{}l@{}}Expert\\ level 3\\ knowledge\\ Prompt\end{tabular}  & \begin{tabular}[c]{@{}l@{}}Prompt 3 = ``can you please extract relevant answer of following question from\\ given content. You are not supposed to write your own answer, according to \\ your understanding, only extract exact lines from relevant content: "\\  text = Prompt 3 + ``{[}" + context{[}i{]} + ``{]}" + ``\textless{}" + question{[}i{]} + ``\textgreater{}"\end{tabular} \\ \hline
\end{tabular}}
\end{table}

\begin{align}
\label{QA-measures}
\left\{
\begin{aligned}
 \text{ROUGE-1} & = \frac{\text{Count of unigrams in ref}}{\text{Count of overlapping unigrams in ref and gen ans}} \\
 \text{ROUGE-2} & = \frac{\text{Count of bigrams in ref}}{\text{Count of overlapping bigrams in ref and gen ans}} \\
 \text{ROUGE-L} & = \frac{\text{Count of words in ref}}{\text{Length of LCS between ref and gen ans}} \\
 \text{ROUGE-S} & = \frac{\text{Count of skip-bigrams in ref}}{\text{Count of overlapping skip-bigrams in ref and gen ans}} \\
 \text{METEOR} & = \text{Precision} \times \left(1 - \beta \times \frac{\text{Insertions}}{\text{Ref Length}}\right) \\
 \text{BLEU} & = \text{BP} \times \exp\left(\sum_{n=1}^{N} w_n \cdot \log(\text{precision}_n)\right) \\
 & \text{Where BP (Brevity Penalty)} & 
\end{aligned}
\right.
\end{align}